\documentclass[iop]{emulateapj}

\usepackage{graphicx}
\usepackage{grffile}
\graphicspath{{figures/}}

\newcommand{\snia}{SN~Ia}
\newcommand{\sneia}{SNe~Ia}
\newcommand{\dmft}{$\Delta M_{15}(B)$}
\newcommand{\cofs}{$^{56}$Co}
\newcommand{\site}{$^{28}$Si}
\newcommand{\nifs}{$^{56}$Ni}
\newcommand{\mch}{$M_{\rm Ch}$}
\newcommand{\supernu}{\textsc{SuperNu}}

\begin{document}

\submitted{Published by ApJ}
\title{Light Curves and Spectra from a Thermonuclear Explosion of a White Dwarf Merger}
\author{Daniel R. van Rossum$^1$, Rahul Kashyap$^{2}$, Robert Fisher$^{2}$, Ryan T. Wollaeger$^{3}$, Enrique Garc\'{i}a-Berro$^{4,5}$, Gabriela Aznar-Sigu\'{a}n$^{4,5}$, Suoqing Ji$^{6}$, Pablo Lor\'{e}n-Aguilar$^{7}$}

\altaffiltext{1}{Flash Center for Computational Science, Department of Astronomy \& Astrophysics, University of Chicago, USA}
\altaffiltext{2}{Department of Physics, University of Massachusetts Dartmouth, 285 Old Westport Road, North Dartmouth, MA 02740, USA}
\altaffiltext{3}{Los Alamos National Lab, NM, USA}
\altaffiltext{4}{Departament de F\'{i}sica, Universitat Polit\`{e}cnica de Catalunya, c/Esteve Terrades, 5, E-08860 Castelldefels, Spain}
\altaffiltext{5}{Institut d'Estudis Espacials de Catalunya, Ed. Nexus-201, c/Gran Capit\`a 2-4, E-08034 Barcelona, Spain}
\altaffiltext{6}{Department of Physics, Broida Hall, University of California Santa Barbara, Santa Barbara, CA 93106-9530, USA}
\altaffiltext{7}{School of Physics, University of Exeter, Stocker Road, Exeter EX4 4QL, UK}


\begin{abstract}

Double-degenerate (DD) mergers of carbon-oxygen white dwarfs have recently emerged as a leading candidate for normal Type Ia supernovae (SNe Ia). However, many outstanding questions surround DD mergers, including the characteristics of their light curves and spectra. We have recently identified a spiral instability in the post-merger phase of DD mergers and demonstrated that this instability self-consistently leads to detonation in some cases.
We call this the spiral merger \snia\ model.
Here, we utilize the \supernu\ radiative transfer software to calculate three-dimensional synthetic light curves and spectra of the spiral merger simulation with a system mass of 2.1 $M_\odot$ from \citet{kashyapetal15}.
Because of their large system masses, both violent and spiral merger light curves are slowly declining.
The spiral merger resembles very slowly declining SNe Ia, including SN 2001ay, and provides a more natural explanation for its observed properties than other \snia\ explosion models.
Previous synthetic light curves and spectra of violent DD mergers demonstrate a strong dependence on viewing angle, which is in conflict with observations.
Here, we demonstrate that the light curves and spectra of the spiral merger are less sensitive to the viewing angle than violent mergers, in closer agreement with observation.
We find that the spatial distribution of \nifs\ and IMEs follows a characteristic hourglass shape.
We discuss the implications of the asymmetric distribution of \nifs\ for the early-time gamma-ray observations of \nifs\ from SN 2014J.
We suggest that DD mergers that agree with the light curves and spectra of normal SNe Ia will likely require a lower system mass.

\end{abstract}

%-----------------------------------------------------------------------
\section{Introduction}\label{sec:Intro}

Type Ia supernovae (SNe Ia) are among the most energetic explosions in the universe. SNe Ia are defined by an absence of hydrogen and strong silicon absorption lines in their spectra. The majority of SNe Ia events, denoted as ``normal'', fall within a relatively narrow range of intrinsic luminosity.
The discovery of a correlation between the intrinsic luminosity and the width of the light curve of SNe Ia, known as the Phillips relation \citep {phillips93}, enabled the use of SNe Ia as standardizable cosmological candles and ushered in a new era of astronomy leading to the discovery of the acceleration of the universe \citep {Riess_1998, Perlmutter_1999}. SNe Ia continue to play a crucial role in the determination of cosmological parameters and serve as leading probes of possible new physics underlying dark energy \citep {weinbergetal13}.

However, despite their central importance to cosmology, the nature of the stellar systems which give rise to SNe Ia remains a long-standing mystery. Unlike the case of core-collapse SNe, whose luminous massive stellar progenitors have been directly observed (most notably in SN 1987A), no stellar progenitor has ever been observed in a normal-brightness SN Ia \citep {maozetal14}.

Isolated white dwarfs (WDs) are inherently stable, and so for many years the leading model for SNe Ia invoked the explosion of a WD in a binary system with a main-sequence or red giant companion. In this \textit{single-degenerate} (SD) channel,  accretion from the companion onto the WD leads to the formation and explosion of Chandrasekhar-mass (\mch) WDs. However, recent mounting observational evidence suggests a diversity of progenitors, including a significant population of sub-Chandrasekhar-mass (sub-\mch) systems \citep {ruiteretal11, scalzoetal14, childressetal15}. As a result, a competing model consisting of the merger of a binary WD system, known as the double-degenerate (DD) channel, has gained increasing attention and may actually account for some if not most normal SNe Ia. The SD and DD scenarios are not the only possible scenarios which might lead to an SN Ia explosion. Additionally, other possible scenarios include the core-degenerate channel \citep {sparksstecher74, livioriess03, kashisoker11, ilkovsoker13, aznarsiguanetal15} and the WD collisional scenario \citep {raskinetal09, rosswogetal09, thompson11, aznarsiguanetal13, kushniretal13}.

It should be noted, however, that recent observations have produced suggestive evidence in support of the SD channel.  The pre-maximum light shock signature of the companion star has been detected in both a subluminous SN Ia \citep {caoetal15} and a normal SN Ia \citep {marionetal15}, although many other searches \citep {ollingetal15} have failed to detect such shock signatures. Moreover, recent X-ray observations of SNRs suggest that both 3C 397 \citep {yamaguchietal15} and Kepler \citep {katsudaetal15} may be of an SD origin. Yet, to date, no convincing stellar ex-companions have been detected in Kepler \citep {kerzendorf_etal_2014}, and the reddening and age of 3C 397 would make such a search particularly challenging.  Consequently, while the case for SD progenitors is building, it has not yet been fully sealed.

In contrast, the DD channel offers natural solutions to a range of outstanding challenges surrounding SNe Ia, including  the absence of H$\alpha$ in the nebular phase \citep {leonard07, shappeeetal13a}, and the delay time distribution (DTD) \citep {totanietal08,  maozbadenes10, maozetal10, grauretal11, maozetal12, grauretal14}.  However, while the DD channel offers natural solutions to many of these outstanding challenges, it nonetheless faces several major hurdles which will need to be overcome before it is likely to be widely accepted as a key contributor to normal SNe Ia. One particularly important hurdle which must be cleared is the demonstration that the spectra and light curves predicted from the DD mergers are in good agreement with the observed spectra and light curves of normal SNe Ia.

A pioneering calculation of the spectra, light curves, and shock breakout signature of an \snia\ from DD models, predating even the first calculations of violent mergers, was made in \citet{fryeretal10}.
In the absence of successful DD SNe Ia models available at that time, \citet{fryeretal10} employed a SD \mch\ gravitationally confined detonation model in their calculations.
It is now understood that the detonation of sub-\mch\ WDs, such as those of the WDs in DD mergers, differ fundamentally from those of \mch\ WDs \citep{simetal10, fisherjumper15}.
Later, synthetic light curves calculated for violent mergers \citep{pakmoretal10, pakmoretal11, pakmoretal12, Moll14} and delayed detonation mergers \citep{Raskin_2014} established that the light curves of DD mergers with total mass exceeding the Chandrasekhar mass would generally be broader than normal SNe Ia.
Moreover, this later work also demonstrated that violent mergers generally have a strong viewing angle dependence and lie off the main Phillips relation \citep{Moll14}.
Synthetic spectra also indicated that mergers posited to arise after disk accretion generally produced very weak intermediate-mass element (IME) absorption features, including Si \textsc{ii} which is a key characteristic of normal SNe Ia \citep {Raskin_2014}.

The majority of more recent theoretical work on the DD channel to date has focused primarily on the violent merger mechanism \citep{pakmoretal10, pakmoretal11, pakmoretal12, danetal12, raskinetal12}.
During the final binary WD merger, the secondary WD is tidally disrupted and rapidly accreted onto the primary over a dynamical timescale, while the primary WD remains relatively intact.
The tidally disrupted secondary then forms a hot, virialized accretion disk surrounding the primary  \citep{Loren_Aguilar_2010, schwabetal12, Zhu_2013, danetal14}.
The violent merger mechanism itself hinges crucially upon a key physical property of the merger.
Specifically, the peak temperature achieved during this tidal disruption process becomes comparable to the carbon ignition temperature for sufficiently massive primary WD masses ($> 1.0 M_{\odot}$) in near-equal mass systems.
In some simulations presented in the literature \citep{pakmoretal10, pakmoretal11, pakmoretal12, Moll14, Raskin_2014}, it is argued that the temperatures and densities achieved in some systems are sufficient to ignite a detonation front according to detonation criteria -- e.g., \citet{Seitenzahl_2009}.
Consequently, in violent merger simulations, a detonation is introduced artificially into the calculation for binary carbon-oxygen (CO) WD systems with a system mass as low as $1.6 M_{\odot}$ \citep{Raskin_2014}.
However, other authors \citep{danetal12, danetal14, raskinetal12} find that pure CO WDs typically only ignite for the very most massive systems ($> 2.1 M_{\odot}$).

A key question one might ask, is whether such DD CO WD systems with a larger system mass become increasingly unstable to the ignition of a carbon detonation front subsequent to the initial merger.
In \citet{kashyapetal15}, we demonstrated both analytically and numerically that binary WD mergers are generally susceptible to a $m = 1$ spiral-mode instability in the hot inner disk produced by the tidal disruption of the secondary.
We further demonstrated through a series of three-dimensional (3D) studies that this $m = 1$ spiral-mode instability transports hot disk material inward, and self-consistently gives rise to a detonation for the case of a $1.1 + 1.0 M_{\odot}$ CO WD binary.
We call this the spiral merger model for SNe Ia.

The primary goal of this paper is to compute the synthetic light curves and spectra from a spiral merger SNe Ia model.
The spiral instability-driven SNe Ia resemble violent mergers in many respects, including their total system mass, mass ratios, and nucleosynthetic yield of \nifs.
However, they differ in one crucial respect, that is, being delayed by about one outer dynamical time from the initial merger, which is posited to be the onset of detonation in the case of violent mergers.
Consequently, spiral merger SNe Ia exhibit a greater degree of axisymmetry than violent mergers, even though both models are quite non-isotropic in polar angle.
As we will see, this greater degree of axisymmetry in the spiral merger model leads to a weaker dependence of the synthetic spectra and light curves with viewing angle than the case of violent mergers.

The recent detection of gamma-rays from SN 2014J was a fundamental advance for supernova science \citep{churazovetal14,Diehl14}.
However, at the same time, the gamma-ray detection from SN 2014J posed a new problem.
Specifically, the detection of the 158 and 812 keV gamma-ray lines from the decay of \nifs\ from 2014J occurred at an unexpectedly early time, i.e., about 20 days after the explosion.
Previous authors invoked a  mechanism by which the accretion of a helium belt onto the primary WD could break the spherical symmetry and generate such an early \nifs\ gamma-ray signature  \citep{Diehl14}.
These gamma-ray observations of SN 2014J provide important new constraints on theoretical models, and so, in addition to computing synthetic optical spectra and light curves, in this paper we also compute synthetic gamma-ray light curves for our model to compare directly against SN 2014J.

The layout of this paper is as follows. In section $\S$\ref{sec:Methods}, we discuss the numerical methodology used to treat the hydrodynamical, nucleosynthetic, and radiative transport evolution. In $\S$\ref {sec:Results}, we present the nucleosynthetic and radiative transport results, including synthetic gamma-ray light curves, for our spiral merger simulation. Finally, in $\S$\ref {sec:Conclusions}, we discuss the relevance of our findings to observational and theoretical work on normal SNe Ia and conclude.

%-----------------------------------------------------------------------
\section{Methods}\label{sec:Methods}

\subsection{Hydrodynamics}

As described in an earlier paper \citep {kashyapetal15}, our hydrodynamical simulations use two distinct numerical methods to model the final evolution of the merger through detonation and into the free-expansion phase. We first employ a smoothed-particle hydrodynamics (SPH) code which is adaptive in a Lagrangian sense and is well-suited to evolving a binary WD during the final merger stage. The end state of the merger is then taken as an initial condition for a Eulerian adaptive mesh refinement (AMR) code, which has excellent shock and detonation tracking capabilities and is ideally suited to following the subsequent evolution of the merged system through detonation and into the free-expansion phase. The SPH calculations begin with a merging CO WD binary system, with equal abundances of carbon and oxygen, with masses $1.1 M_{\odot}$ + $1.0 M_{\odot}$, and a mass resolution of $2\times 10^5$ particles. These simulations employed an SPH code described previously in the literature \citep{Loren_Aguilar_2010}. The final merger proceeds on a dynamical timescale as the secondary WD is tidally disrupted and rapidly accreted onto the primary, which remains relatively intact. The results are generally in good agreement with a range of previous studies -- e.g. \citet {Loren_Aguilar_2010}, \citet {schwabetal12} and \citet {danetal14}.

After obtaining the outcome of the merger in SPH,  we subsequently map the SPH data onto a 3D Eulerian grid in FLASH 4.1, which is an AMR grid-based code employing higher-order unsplit Godunov hydrodynamics solvers \citep{Fryxell_2000, Lee_2009}. The SPH data is remapped roughly 40 s, or 1.5 outer rotational periods, subsequent to tidal disruption of the secondary when the maximum temperature is achieved. Our grid domain extends from $-2.8\times 10^{10} $cm to $+2.8\times 10^{10} $cm in each direction.  The AMR simulations principally utilize a mass-based resolution criterion and ensure that no cell has a mass exceeding $4\times 10^{27} $g at any point in space or time. We employ the Helmholtz equation of state, which includes contributions from electrons (including an arbitrary degree of relativity and degeneracy), nuclei, and photons assuming that all are in local thermodynamic equilibrium \citep{Timmes_2000}. We also include nuclear energy generation using a 19 isotope $\alpha$-chain reaction network including the effect of neutrino cooling \citep{Timmes99}. This small nuclear network suffices to capture the dominant  nuclear reactions responsible for heating the gas. A larger network is employed during post-processing to calculate more detailed nucleosynthetic yields (see below). FLASH incorporates both multigrid and multipole solvers to model self-gravity. Here, we employ an improved multipole solver \citep{Couch_2013} with isolated boundary conditions and terms through $\ell = 60$ in the multipole expansion. We have previously demonstrated excellent agreement between multipole and multigrid solvers for the merger simulation discussed in this paper \citep{kashyapetal15}.

In addition to the Eulerian grid-based data, the FLASH simulations also included passive Lagrangian tracer particles, which are employed in computing the detailed nucleosynthetic yields using a large nuclear network and radiative transfer in post-processing. A total of $10^5$ particles are distributed proportional to the gas grid density, effectively tracing the mass. The fluid quantities, including velocity, are interpolated at every timestep from the Eulerian mesh using a quadratic interpolation scheme. The particle positions are in turn integrated using a two-stage, second-order Runge-Kutta method also known as the Heunn method.

The nuclear energy released by the detonation wave accelerates the material to high velocities.
During this expansion, the gravitational deceleration gradually decreases and the velocities approach their asymptotic free-expansion value.
This tendency for velocities to approach their free-expansion values can be illustrated by noting that at a given time after detonation, the relative deviation in expansion velocities $\Delta v / v$ with respect to the asymptotic free-expansion velocity $v$ can be estimated by assuming that the pressure gradients are small, and therefore
\begin{equation}
 \frac{\Delta v}{v} = \frac{1}{v} \int^\infty_t \frac{G M_v}{r(t')^2} d t' \approx \frac{G M_v}{v^3 t} \propto \frac{\bar{\rho_v}}{t} ,
\end{equation}
where $G$ is the gravitational constant, $M_v$, is the mass inside a sphere of radius $r = vt$, and $\bar{\rho_v}$ is the mean density of the sphere in velocity space.
The approximation $v(t) \approx v$ used in the last step underestimates the deviation but is reasonably good if $\Delta v/v$ is small, and demonstrates that the relative deviation $\Delta v / v \rightarrow 0$ as $t^{-1}$.
We follow the deceleration phase in the full hydrodynamic simulation until the deviation in velocities is smaller than 7\% in all regions of the domain.
The regions close to the center take approximately 15 s to reach this condition, which is longer than the outer regions.
This is because the mean density $\bar{\rho_v}$ of a sphere with radius $r = vt$ decreases outward due to the strong negative gradient in the density profile of the ejecta.

In this paper, we define $t = 0$ to be the onset of detonation.

\subsection{Nucleosynthesis}
After the hydrodynamic simulation, we post-process the tracer particles using the TORCH nuclear network code \citep{Timmes99,Townsley15} to calculate the nuclear evolution of a set of 225 nuclides on the temperature-density histories of each of the tracer particles.
In \citet{kashyapetal15}, we considered a single physical $1.1 M_{\odot} + 1.0 M_{\odot}$ CO WD merger system, varying the spatial and temporal resolution, as well as gravity solver.
In this paper, we utilize the data from the ``2-mp2'' run from \citet{kashyapetal15}, which had a maximum finest spatial resolution of 136 km, a multipole solver for gravity, and a further restriction on the timestep set by nuclear burning in addition to the standard Courant-Friedrichs-Lewy (CFL) criterion.
We assume a homogeneous initial composition of 50\% $^{12}$C, 47.1\% $^{16}$O, 2.37\% $^{22}$Ne, and other solar metals with atomic mass number $A>16$ using solar abundances from~\citet{Anders89}\footnote{%
We note that the initial composition used in the FLASH simulation was 50\% $^{12}$C and 50\% $^{18}$O.
We have verified that the difference in energy release between this composition and the composition used in the nucleosynthesis is on the order of a few percent.}.
$^{22}$Ne is produced during core helium burning by helium captures on $^{14}$N left from hydrogen burning via the CNO cycle \citep{Bildsten01}.
The amount of $^{22}$Ne is chosen such that the total electron fraction $Y_e = 0.49886$, which corresponds to a solar metallicity zero-age main-sequence star (see also~\citet{Seitenzahl13}).
\citet{Bildsten01} point out that the spatial distribution of $^{22}$Ne becomes stratified over time in WD interiors due to gravitational sedimentation.
\citet{bravoetal11} considered the effect of $^{22}$Ne sedimentation and found that it did not have an appreciable effect on the observable properties of SNe Ia.
Moreover, the sedimentation time is quite long even for high-mass WDs ($> 10$ Gyr at the half-mass radius for a $1.2\ M_{\odot}$ WD; \citep{Althaus10}), and so we neglect the stratification of $^{22}$Ne here and assume it is uniformly spatially distributed.

The nuclear burning, as computed over the entire duration of our hydrodynamic evolution, determines the isotopic mass fractions in the fluid parcels as represented by each of the particles at the time the hydrodynamic simulation ends.
The nuclear products include a range of radioactive isotopes.
The radioactive $\alpha$-chain isotopes \nifs, $^{52}$Fe, and $^{48}$Cr are produced in large quantities compared to other radioactive isotopes and have half-lives ranging from hours to days, which are comparable to the time scale on which the luminosity reaches its peak.
Therefore, the opacities of the ejecta change significantly on a time scale of hours to weeks due to the decay of these radioisotopes, and the decay chains of these radioisotopes are followed explicitly during the radiation transport simulation.
All other radioactive isotopes are produced in smaller abundances and are approximated to either fully decay for half-lives shorter than 5 days or to not decay at all for half-lives longer than 5 days.

\subsection{Radiation Transport}
Following the hydrodynamic simulation and nucleosynthesis post processing we calculate the light curves and spectra from the ejecta using the \supernu\ software \citep{Wollaeger13,Wollaeger14,Vanrossum15}.
It uses Implicit Monte Carlo (IMC) and Discrete Diffusion Monte Carlo (DDMC) methods to stochastically solve the special-relativistic radiative transport and diffusion equations to order $v/c$ in three dimensions.
The hybrid of IMC and DDMC excels in solving radiation transport problems that involve a large dynamic range of matter densities, so that \supernu\ lends itself well to supernova simulations.
Codes based on traditional Monte Carlo methods or IMC alone are computationally inefficient in regions with high optical depth, so that in supernova simulations it is necessary to make approximations to estimate the rates of energy diffusion, which is one of the key properties of supernova light curves.

\supernu\ follows the free-expansion phase of the supernova using a velocity grid.
Therefore, we map the final velocities (at $t=16.5$\,s) of the tracer particles obtained during the hydrodynamic simulations to a 3D cartesian velocity grid, or a 3D cylindrical velocity grid in this case of the spiral merger which features a significant degree of azimuthal symmetry.
The mass in each of the grid cells equals the sum over the masses of all the particles that are mapped into that cell; the chemical composition equals the mass-weighted average of the chemical composition of those same particles.
Unlike the Eulerian-based FLASH code, \supernu\ handles true vacua within cells; consequently, zero-mass grid cells into which no particles are mapped (in the corners of the 3D cylindrical domain) require no special treatment, and indeed, cells in the radiation transport grid that are not sampled by any particles are treated as vacuum.

In each timestep, radiation that escapes the domain is binned into 600 logarithmically spaced wavelength bins in the range $[10^3, 10^{4.5}]$\AA\ and 9 polar viewing angle bins, linearly spaced in $\mu = \cos(\theta)$ (so that all bins have a fixed amount of solid angle).
These synthetic spectra are subsequently folded over the Bessell band filter functions \citep{Bessell90} to obtain synthetic light curves in the UBVRI bands.

%-----------------------------------------------------------------------
\section{Results}\label{sec:Results}

\subsection{Ejecta Morphology and Nucleosynthetic Yields}

The spiral-mode detonation arises during the accretion (onto the primary) of a spiral arm in the accretion disk that is formed during the disruption of the secondary.
During this accretion, hot, low-density material from the secondary mixes with cold, high-density fuel from the primary, creating a mixture that satisfies detonation criteria \citep[and references therein]{kashyapetal15}.

\begin{figure*}
\centerline{\includegraphics[width=.8\textwidth]{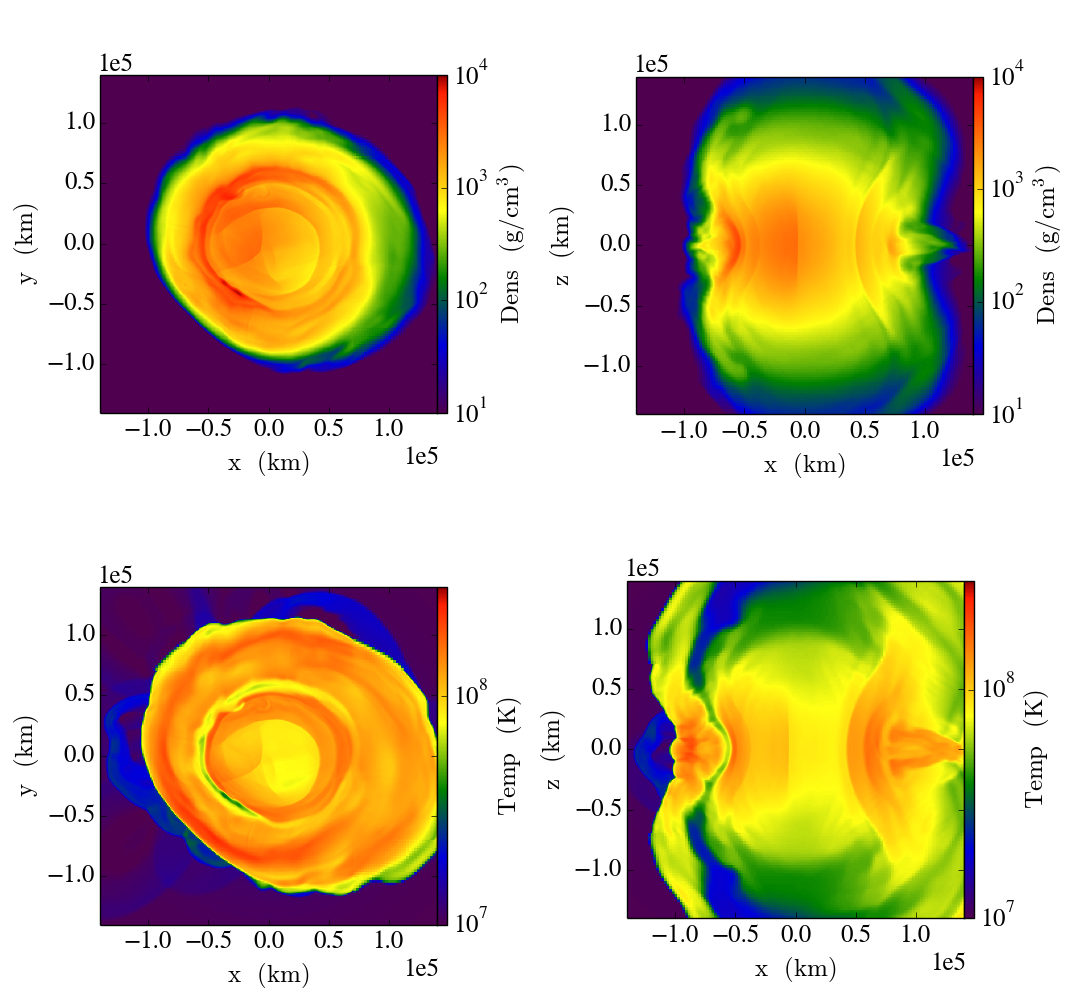}}\vspace{-4pt}
\caption{Four projections of the density and temperature structure of the ejecta (at $t = 12$\,s) in the 3D explosion simulation: on the $xy$-plane (left), $xz$-plane (right).}
\label{fig:lastslice}
\end{figure*}

The presence of a tidally disrupted disk at the time of detonation leaves an imprint on the distribution of material in the ejecta.
The detonation front leaves a substantial amount of unburned carbon and oxygen within the disk and produces almost no iron-peak elements, as revealed by previous simulations of violent mergers \citep{Raskin_2014} and artificially ignited ``tamped'' detonations \citep{Moll14}.
In Figure~\ref{fig:lastslice}, we show the density and temperature structure of the ejecta of the supernova explosion at $t = 12$ s, close to the final time of the hydrodynamic simulation.
The ejecta show a high degree of rotational symmetry along the $z$-axis, as well as reflection symmetry across the $z=0$ plane.
This degree of symmetry is already present before the onset of the detonation.
It is a characteristic feature of the spiral merger model.
The delay to detonation of a few dynamical times gives the system time to reduce asymmetries.
The detonation subsequently does not significantly disturb the degree of symmetry because it propagates faster than the sound speed, and so the material has no time to react to the change in pressures.
Note that the coordinate origin is located at the center-of-mass of the system, so that the primary WD is slightly off-centered on this and other figures.

\begin{figure}
\centerline{\includegraphics[width=.4\textwidth]{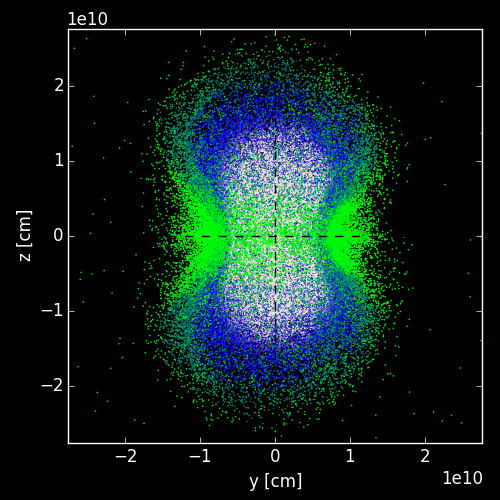}}\vspace{-4pt}
\centerline{\includegraphics[width=.4\textwidth]{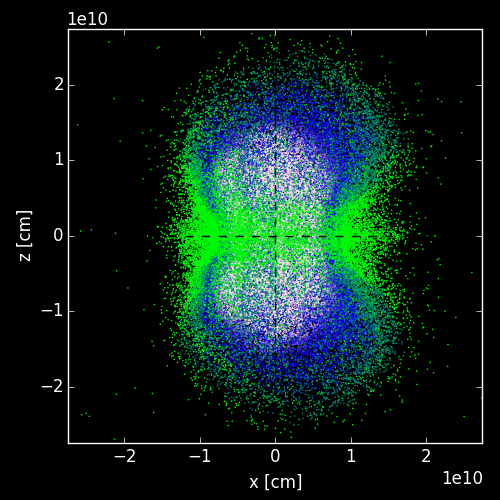}}\vspace{-4pt}
\centerline{\includegraphics[width=.4\textwidth]{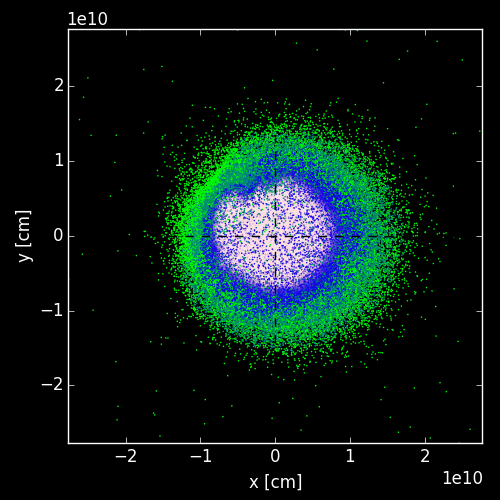}}
\caption{Three projections of the Lagrangian tracer particles at their final positions (at $t = 16.5$\,s) in the 3D explosion simulation: on the $yz$-plane (top), $xz$-plane (middle), and $xy$-plane (bottom).
 The chemical composition of each particle is color coded as unburnt carbon and oxygen (green), intermediate-mass elements (blue), stable iron-peak elements (red), and \nifs\ (white).
 In this explosion, only a small amount of stable iron-peak elements is produced, mainly in the central region together with \nifs.
}\label{fig:particles}
\end{figure}

The distribution of the nucleosynthetic yields of the spiral merger model in the 3D computational domain is shown in Figure~\ref{fig:particles} by color coded tracer particles, projected with their final positions along the $x$, $y$, and $z$-axes in each of the three panels.

Each particle has a mix of green, blue, red, and white color, depending on the chemical composition of the parcel of material that the particle represents.
The white color represents \nifs, and the green, blue, and red colors in the figure represent isotopes with $A \leq 16$, intermediate-mass isotopes with $16 < A \leq 40$, and iron-peak isotopes for all $A > 40$ except \nifs, respectively\footnote{%
We note that this definition of colors yields unique colors for arbitrary abundances because the four groups of abundances are normalized, i.e.\ the sum over the four groups is one.
White acts as a brightness channel.
The primary colors red, green, and blue together can yield a dark gray color if all of the respective abundances are equal, e.g.\ three times 1/3.
However, this yields a different color than a cell that contains \nifs\ because the white channel increases the brightness three times faster than a combination of the primary colors.}.
The fact that iron-peak elements apart from \nifs, depicted in red, are not very prevalent indicates that the iron-peak products are mainly \nifs, as one expects from the sub-\mch primary \citep {seitenzahletal13, seitenzahletal15}.
Table~\ref{tab:yields} lists the nucleosynthetic yields of the explosion simulation, limited to the abundant $\alpha$-chain isotopes out of the 225 isotopes that are followed in the network calculation, and the final decay products of iron-group isotopes ($A > 40$).
About 14\% of the iron-group consists of isotopes other than \nifs.
\begin{deluxetable}{lc|lc}
\tablecaption{Nucleosynthetic Yields and Final Decay Products\label{tab:yields}}
\tablehead{\colhead{Isotope} & \colhead{Yield} & \colhead{Decay} & \colhead{Yield}\\
                       & \colhead{[$M_\odot$]$^a$} & \colhead{Product} & \colhead{[$M_\odot$]$^a$}}
\tablenote{Powers of 10 are written in parentheses.}
\startdata
$^{ 4}$He & 5.82(-3) &   Ti & 3.21(-4)\\
$^{12}$C  & 0.300    &    V & 6.77(-5)\\
$^{16}$O  & 0.499    &   Cr & 7.02(-3)\\
$^{20}$Ne & 2.05(-2) &   Mn & 4.57(-3)\\
$^{24}$Mg & 4.66(-2) &   Fe & 0.678   \\
$^{28}$Si & 0.318    &   Co & 3.23(-5)\\
$^{32}$S  & 0.112    &   Ni & 4.45(-2)\\
$^{36}$Ar & 1.71(-2) &   Cu & 3.10(-5)\\
$^{40}$Ca & 1.43(-2) &   Zn & 1.49(-4)\\
$^{44}$Ti & 1.49(-5) \\
$^{48}$Cr & 2.79(-4) \\
$^{52}$Fe & 6.37(-3) \\
$^{56}$Ni & 0.629    \\
$^{60}$Zn & 8.99(-3) \\
\hline
$^{12}$C + $^{16}$O & 0.800 & \phantom{\large X} \\
$16 < A \leq 40$ & 0.571 \\
$A$ $>$ 40 & 0.729 \\
\hline
Total & 2.10 & \phantom{\large X}
%\hline
%$E_{\rm kin}$[erg] & 1.07 $\times 10^{51}$
\enddata
\end{deluxetable}%

We also note that our nucleosynthetic yields are well converged even at relatively modest particle counts. We randomly downsampled our final particles and recomputed our yields on the downsampled particle distribution. We further carefully verified that (as expected) the downsampled particles track the same underlying mass distribution as the original distribution using the KS test, and so this downselection process is essentially the same as beginning a new calculation with the lower particle count. We find that the abundances are very well converged (to within 0.2\%) for $10^5$ particles for major species. Remarkably, even for total particle counts as low as $10^3$, the key abundances are still converged to within $\approx$ 3 - 4\%.

Earlier work by other authors \citep {seitenzahletal10} demonstrated that relatively high counts ($> 10^5$) of Lagrangian tracer particles uniformly distributed by mass were required in order to achieve convergence in Chandrasekhar-mass, single-degenerate models. However, owing to their significantly higher central concentrations, these Chandrasekhar-mass WDs necessarily require much higher particle counts than sub-Chandrasekhar-mass models, such as those studied here, to achieve good convergence (see Table \ref{tab:yieldres}).

In order to elaborate upon this issue of convergence in particle count, consider the Poisson statistical distribution of particles within a mass range $\Delta M$. While many physical and numerical parameters enter into the overall error, the particle distribution in mass may be thought of as a random process over the particle number, and consequently shot noise dominates the overall error budget at low particle counts typical of numerical simulations. Specifically, Poisson statistics imply a standard deviation in mass-weighted abundances $\sim \sqrt {M / \Delta M} \sqrt {1 / N}$, where $N$ and $M$ are the total number of particles and the total mass of the system, respectively, and $\Delta M$ is a given mass range. The first factor of $\sqrt {M / \Delta M}$ simply depends on a given mass profile within a given mass range $\Delta M$, and the second factor of $\sqrt {1 / N}$ simply depends on the total particle count $N$. Consequently,  the relative fractional errors within some mass range $\Delta M$ for any model, whether Chandrasekhar-mass or sub-Chandrasekhar, at a fixed particle count is dependent solely upon $\sqrt {M / \Delta M}$.

We may very roughly estimate the impact of Poisson statistics on the nucleosynthetic yields of the IME by considering the tail of the cumulative mass distribution below densities of $10^7$ g/cm$^{-3}$. In this case, the factor of $(M / \Delta M)^{1/2}$ is 7.78 for a Chandrasekhar-mass WD and 1.27 for the 1.1 + 1.0 C/O WD system in this paper. Put differently, we estimate that the particle resolution requirements for IMEs is some $(7.78 / 1.27)^2 \sim 40$ times higher for the aforementioned Chandrasekhar-mass model than the sub-Chandrasekhar-mass merger model in the paper. This estimate is in good agreement with \citet{seitenzahletal10}'s findings that even a $32^3$ model is converged to within $\approx$5\%, which is close to the error levels for our downsampled $10^3$ ($\sim 32^3 / 40$) distribution.

\begin{deluxetable}{l|cccc}
\tablecaption{Nucleosynthetic Yield Convergence with Particle Count\label{tab:yieldres}}
\tablecomments{All abundances given in solar masses.}
\tablehead{\colhead{Particle Number} & \colhead{$^{56}$Ni} & \colhead{$^{28}$Si} & \colhead{$^{40}$Ca} & \colhead{$^{32}$S }}
\startdata
$10^5$ & 0.629 & 0.327 & 0.014 & 0.116 \\
$2.5 \times 10^5$ & 0.630 & 0.327 & 0.014 & 0.116 \\
$10^4$ & 0.639 & 0.317 & 0.013 & 0.110 \\
$10^3$ & 0.615 & 0.316 & 0.015 & 0.111
\enddata
\end{deluxetable}%

\begin{figure}
\centerline{\includegraphics[width=.45\textwidth]{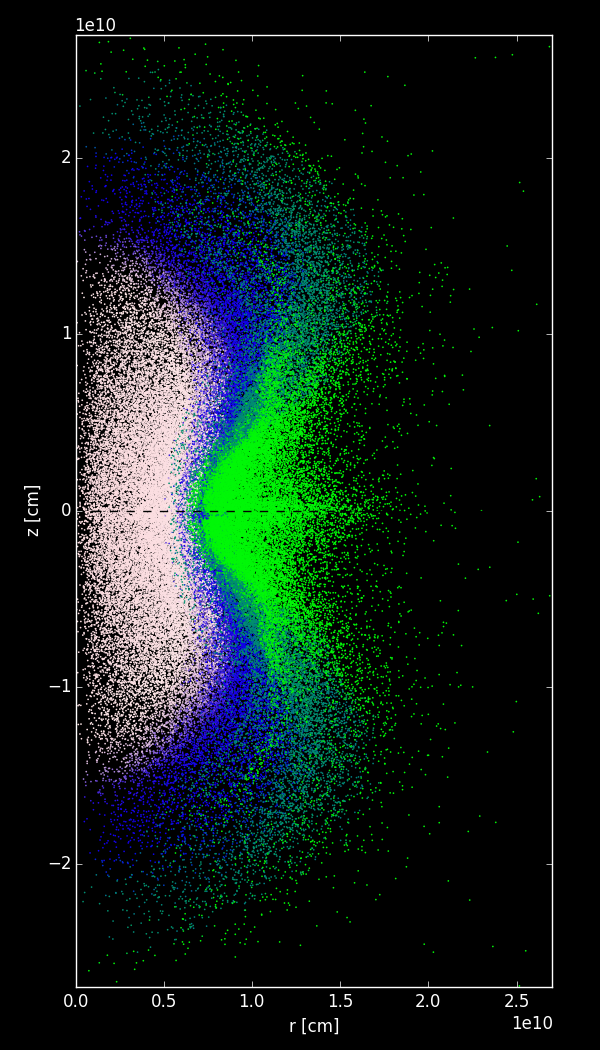}}
\caption{Lagrangian tracer particles color coded in the same way as in Figure~\ref{fig:particles}, with their final positions (at $t = 16.5$\,s) in the explosion simulation mapped to cylindrical geometry and azimuthally projected onto the $\phi=0$ plane.
 This demonstrates the high degree of reflection symmetry across the $z=0$ plane.
 The radiation transport calculations with SuperNu are performed in 3D cylindrical geometry, exploiting the considerable degree of axisymmetry in the ejecta.
Note that in cylindrical geometry, the volume of cylindrical shells around the symmetry axis is proportional to the radial distance.
Consequently, the particle number density close to the symmetry axis appears lower than at slightly larger distances from the axis, even though the particle number density is proportional to the material density everywhere.
}\label{fig:particles-cylinder}
\end{figure}

Figure~\ref{fig:particles-cylinder} shows the distribution of tracer particles with their final positions mapped to cylindrical geometry and azimuthally projected onto the $\phi=0$ plane.
This demonstrates the degree of reflection symmetry across the $z=0$ plane.

The velocity scale over which the isotopic stratification changes in the ejecta -- from iron-group elements close to the center, to intermediate-mass elements at higher velocities, and finally to unburned carbon and oxygen at the highest velocities -- strongly depends on polar angle.
At small inclination angles from the plane of the disk, the ejecta are significantly slower (see the quantitative discussion below) than at higher inclination angles, similar to \citet{Raskin_2014}.
This is caused by momentum conservation when the accelerated material sweeps up the mass of the accretion disk, which initially has a radial velocity close to zero.
At higher inclination angles, the amount of mass to be swept up decreases and the radial velocity that is reached in the ejecta is higher.

\begin{figure}
\centerline{\includegraphics[width=.5\textwidth]{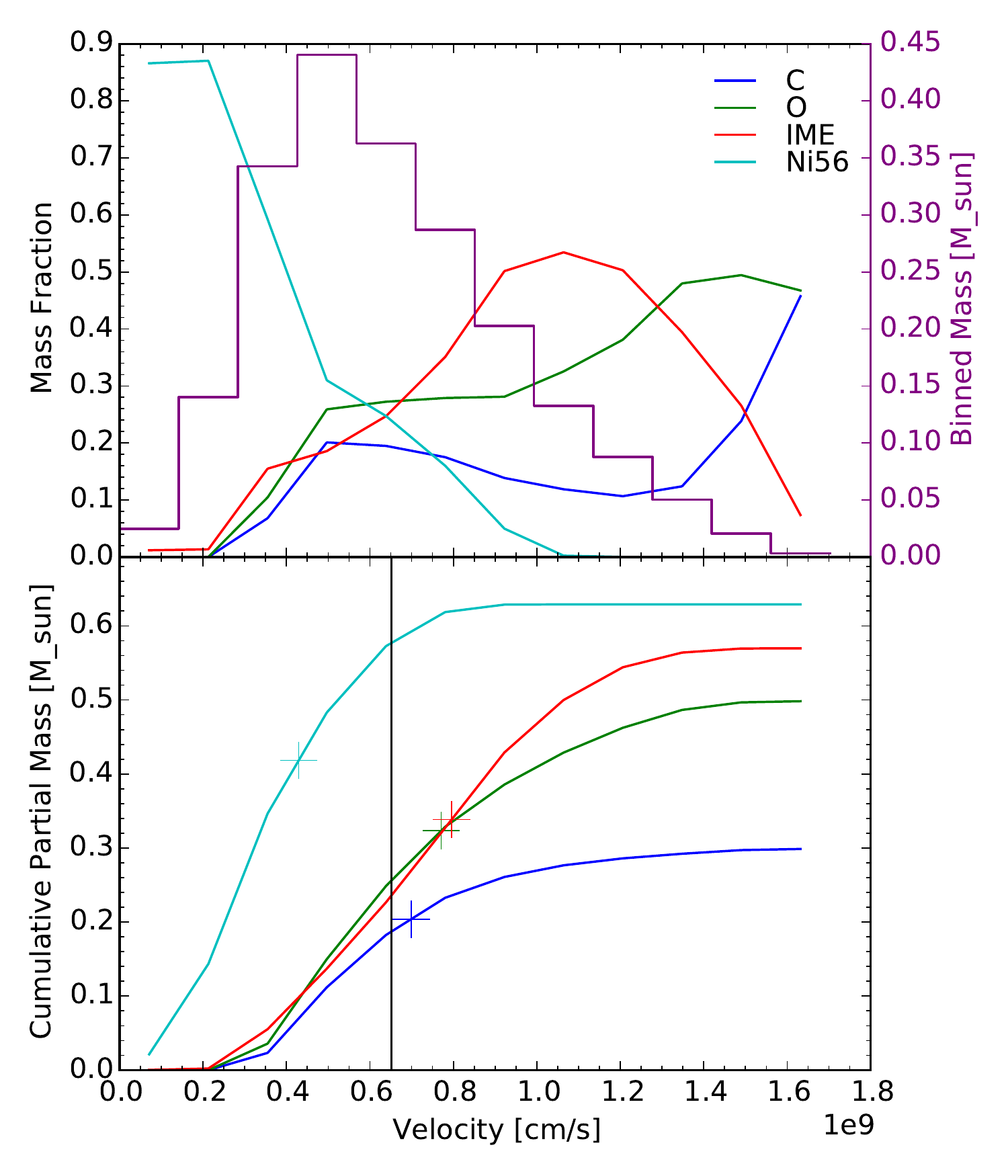}}
\caption{Top panel: Mass fractions of different chemical elements in bins of the ejecta velocity.
 Elements with atomic number $A \leq 40$, except carbon and oxygen, are grouped together as IME\@.
 The total mass in each velocity bin is plotted with a purple line (right-hand axis).
 Bottom panel: Cumulative partial masses of the same chemical elements.
 The respective mass-weighted average velocities are indicated with crosses.
 The mass-weighted average velocity of the total ejecta, $6.5 \times 10^8$ cm/s, is indicated with a vertical black line.
}\label{fig:profile1Dsph}
\end{figure}

The expansion velocities of the ejecta, or the kinetic energy, play an important role in the shape of the bolometric light curve.
Higher expansion velocities let the densities decrease more quickly, so that the energy from radioactive decay that is deposited in the core can leak out sooner, causing the light curves to peak earlier.
Figure~\ref{fig:profile1Dsph} shows the mass distribution profile on a radial velocity scale together with abundance profiles of carbon, oxygen, intermediate-mass elements (IMEs), and \nifs\ in the top panel, similar to \citet[Figure 5]{Moll14}.
In the bottom panel, cumulative mass fractions are plotted on the same radial velocity scale.
The crosses on each of the cumulative mass fraction curves indicate the mass-weighted average velocity of the respective materials in the ejecta: $4.29 \times 10^{8}$ cm/s for \nifs, $6.99 \times 10^{8}$ cm/s for carbon, $7.71 \times 10^{8}$ cm/s for oxygen, and $7.96 \times 10^{8}$ cm/s for intermediate-mass elements.
A dotted vertical line at $6.52 \times 10^{8}$ cm/s indicates the mass-weighted average velocity of all of the ejecta.
These numbers are important in conjunction with the width of light curves and can be compared against the results in the literature.

\begin{figure}
\centerline{\includegraphics[width=.5\textwidth]{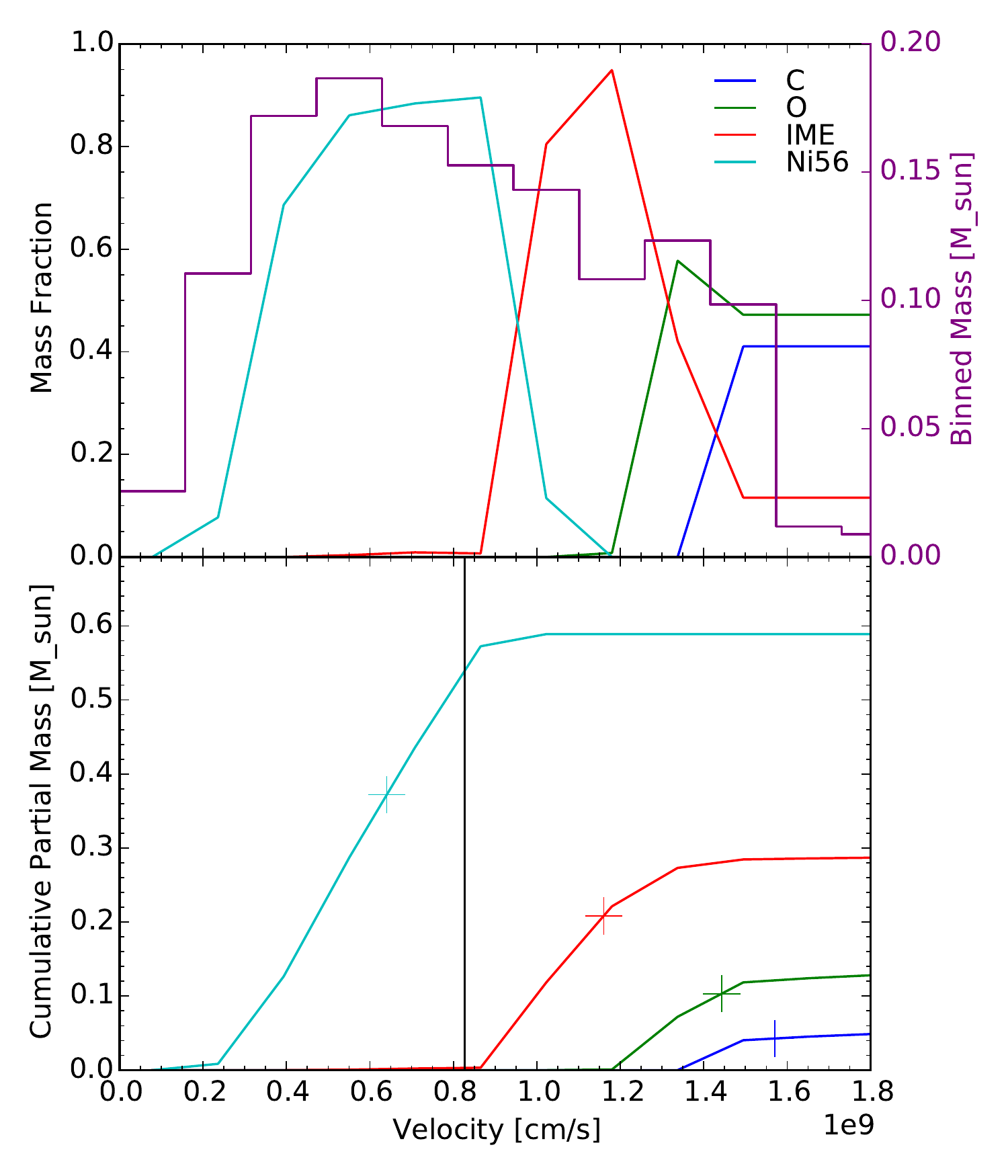}}
\caption{Same as Figure\,\ref{fig:profile1Dsph} but for the W7 model.
 The W7 model has a total mass of 1.32 $M_\odot$, which is 35\% lower than the 2.1 $M_\odot$ spiral merger simulation.
 The mass-weighted average velocity of the total ejecta is $8.3 \times 10^{8}$ cm/s.
}\label{fig:w7-profile1Dsph}
\end{figure}

The mass distribution of the classical W7 mode \citep{Nomoto84} is plotted in Figure\,\ref{fig:w7-profile1Dsph} for comparison.
W7 is a one-dimensional phenomenological \mch\ \snia\ model.
The \nifs\ mass of the W7 model is approximately the same as the spiral merger simulation and the mass-weighted average velocity is 28\% higher.

\begin{figure}
\centerline{\includegraphics[width=.5\textwidth]{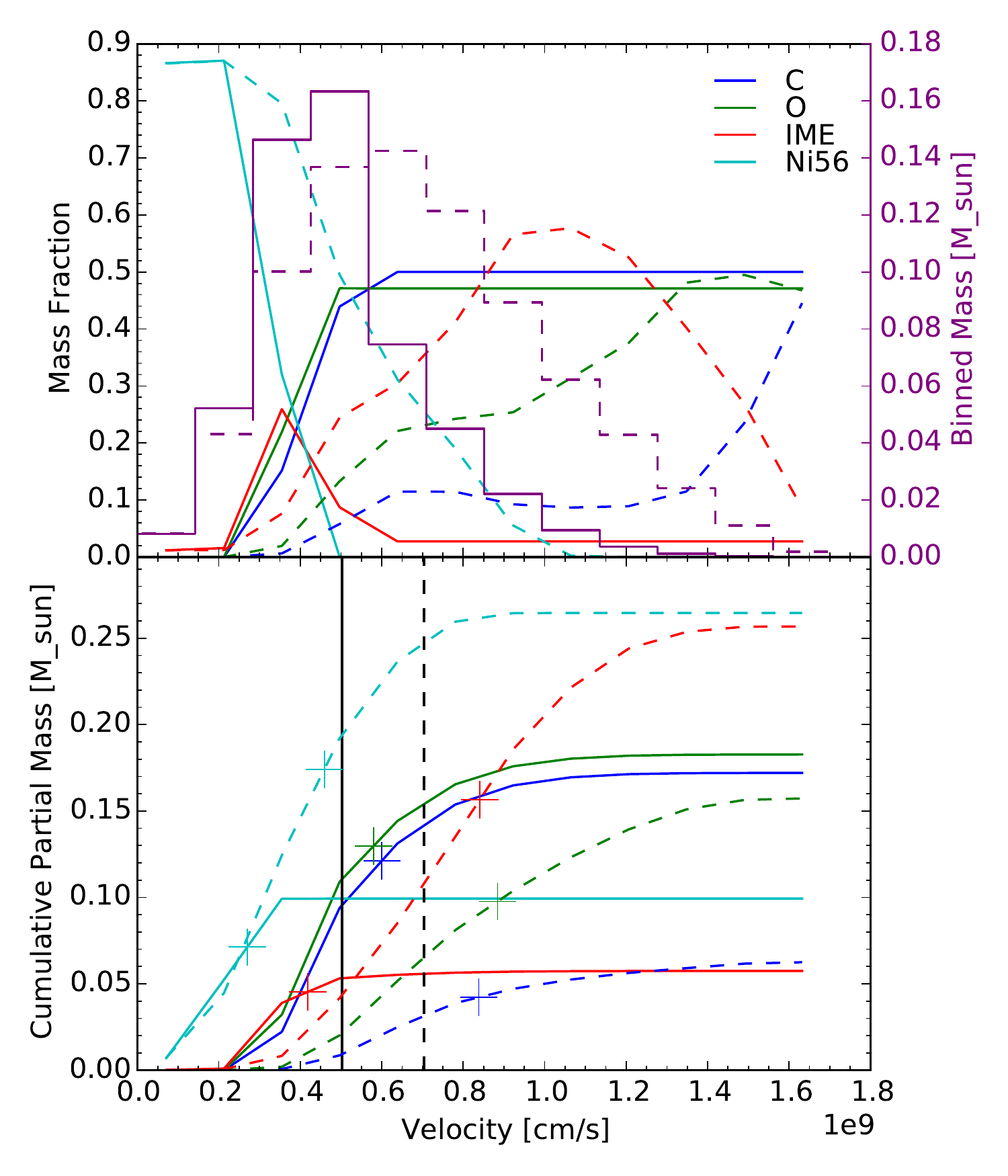}}
\caption{Same as Figure~\ref{fig:profile1Dsph} but with the ejecta split over velocity polar angles.
 The central region $-0.33 \leq \mu = \cos(\theta) < 0.33$, which includes the $z=0$ plane, is plotted with solid lines.
 The $\mu > 0.33$ cone that includes the north pole is plotted with dashed lines.
 The $\mu < 0.33$ cone is very similar to its northern counterpart and is omitted to simplify the plot.
 The total mass in the central bin is $0.53\ M_\odot$ and $0.78\ M_\odot$ in each of the polar bins.
 The mass-weighted average velocity of the ejecta is $5.0 \times 10^8$ cm/s around the $z=0$ plane and $7.0 \times 10^8$ cm/s toward the poles.
}\label{fig:profile2Dsph}
\end{figure}

The polar angle asymmetry in the ejecta that is caused by the presence of the spiral disk at the time of detonation can be quantified using angle-dependent profiles similar to the spherical profiles of Figure~\ref{fig:profile1Dsph}.
Figure~\ref{fig:profile2Dsph} shows these profiles for several velocity polar angle bins.
Three bins are used and are regularly spaced in $\mu = \cos(\theta)$ so that the volume of the velocity bins is constant with polar angle.
Only the central bin and the northern bin are plotted because the profiles in the two opposite polar bins are very similar.
The plot shows that the mass-weighted velocity is on average $7.0/5.0 = 1.4$ times higher toward the poles than in the bin that contains the spiral disk.
Apparently, \nifs\ is distributed spherically out to a velocity of $3 \times 10^8$ cm/s, but almost exclusively in the polar bins beyond that velocity.
Approximately 11\% of the elements with atomic number $A \leq 40$ except carbon and oxygen, together referred to as IME, are located in the central polar bin and their mass-weighted velocity is slower by a factor of $4.2 / 8.4 = 0.50$.

\begin{figure*}
\centerline{\includegraphics[trim=40 65 20  0, width=.5\textwidth,clip]{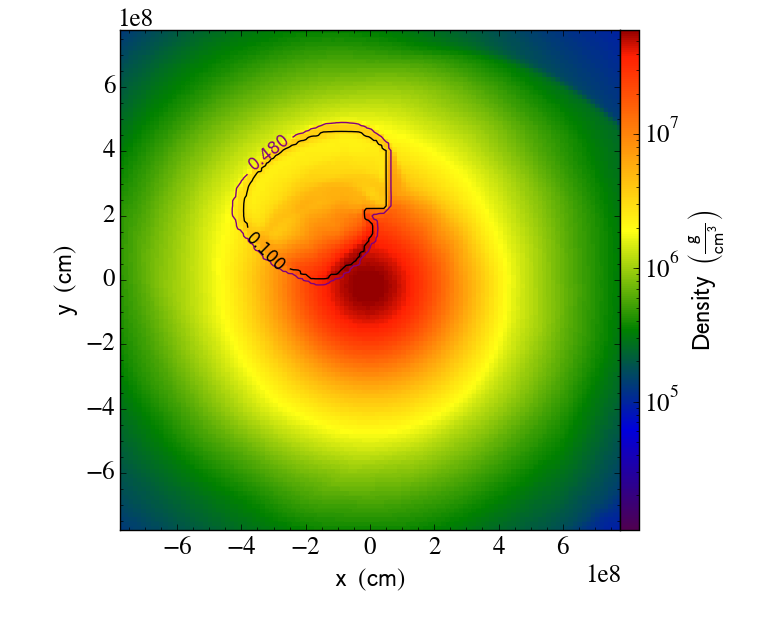}
            \includegraphics[trim=40 65 20  0, width=.5\textwidth,clip]{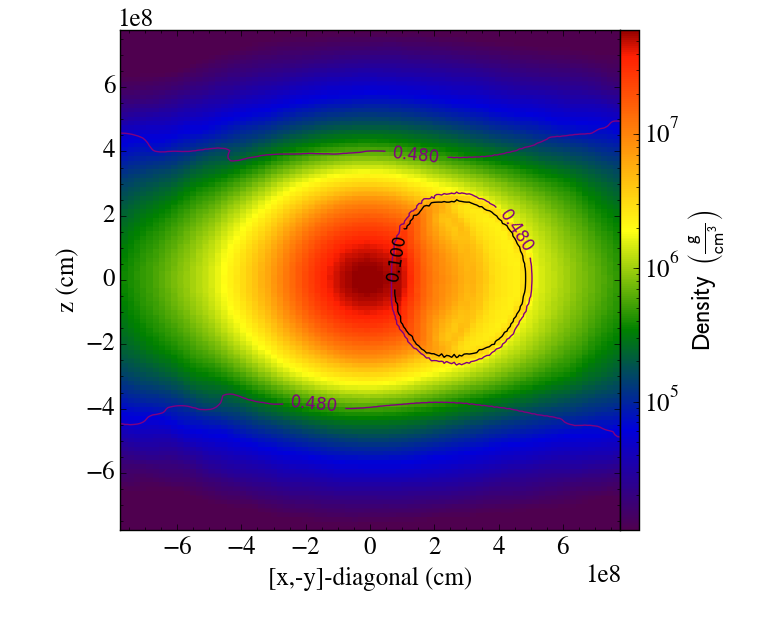}}
\vspace{-1pt}
\centerline{\includegraphics[trim=40 20 20 22, width=.5\textwidth,clip]{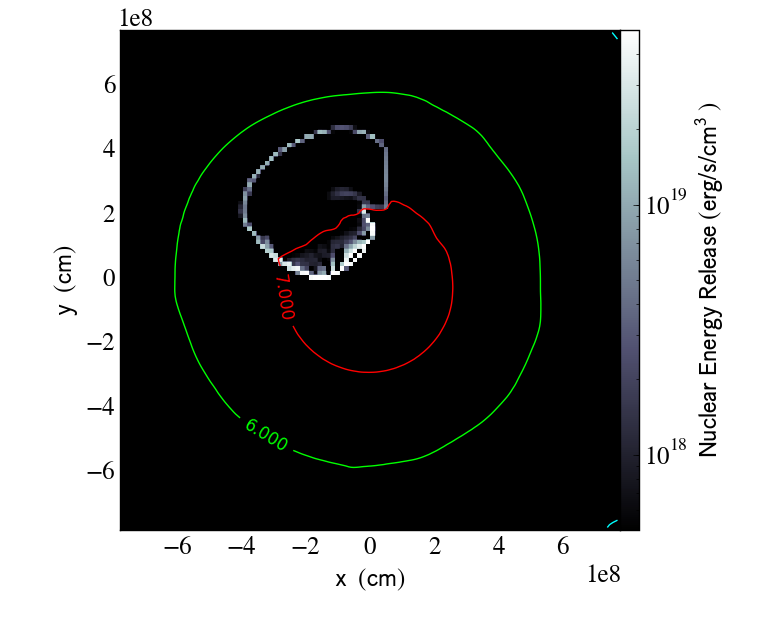}
            \includegraphics[trim=40 20 20 22, width=.5\textwidth,clip]{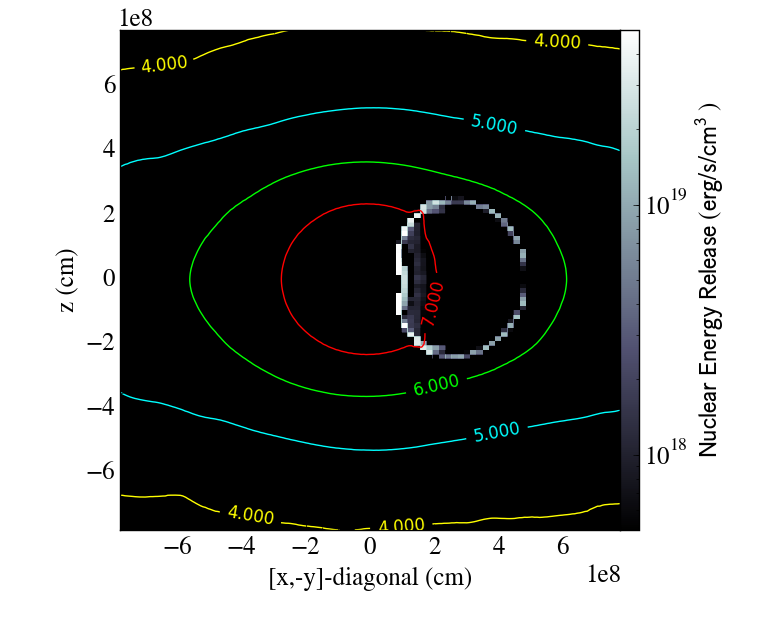}}
\caption{Density color-map with contours of the $^{12}$C mass fraction (top row) and nuclear energy release rate with density contours ($\log_{10}(\rho) \in [4,5,6,7]$, bottom row) in the $z=0$ plane (left column) and the plane spanned by the $z$-axis and the $(x,y) = (1,-1)$ vector (right column) at $t=0.34$ s.
 The $^{12}$C mass fraction contours at values lower than 0.5, which is the initial mass fraction for $^{12}$C, show where material has been processed by the detonation wave.
 The density color-map shows the location of the primary ($\log_{10}(\rho) \gtrsim 7$, red contour), and the surrounding disk.
 The energy release rate at $t=0.34$ s is large in the disk, as well as in the primary.
 The energy release inside the disk gives rise to the relatively strong vertical acceleration of the disk material and redistributes that material away from the $z=0$ plane.
}\label{fig:diskdetonation}
\end{figure*}

The fraction of mass in the central bin, $0.53/2.1 = 0.25$, is 35\% lower than the fraction in the polar bins, $0.78/2.1 = 0.38$.
This is interesting because it is the opposite of the initial configuration where approximately 1/3 of the primary mass and all of the secondary mass is located in the central bin.
When the detonation propagates through the disk -- where it eventually quenches due to low densities -- the large vertical density gradient and small horizontal density gradient give rise to predominantly vertical acceleration.
This is demonstrated in Figure~\ref{fig:diskdetonation}, which shows the detonation wave, at $t=0.34$ s, burning $^{12}$C and releasing energy in a region that extends from the primary into the disk.

%-----------------------------------------------------------------------
\subsection{Synthetic Light Curves}\label{sec:lc}

\begin{figure}
\centerline{\includegraphics[width=.5\textwidth]{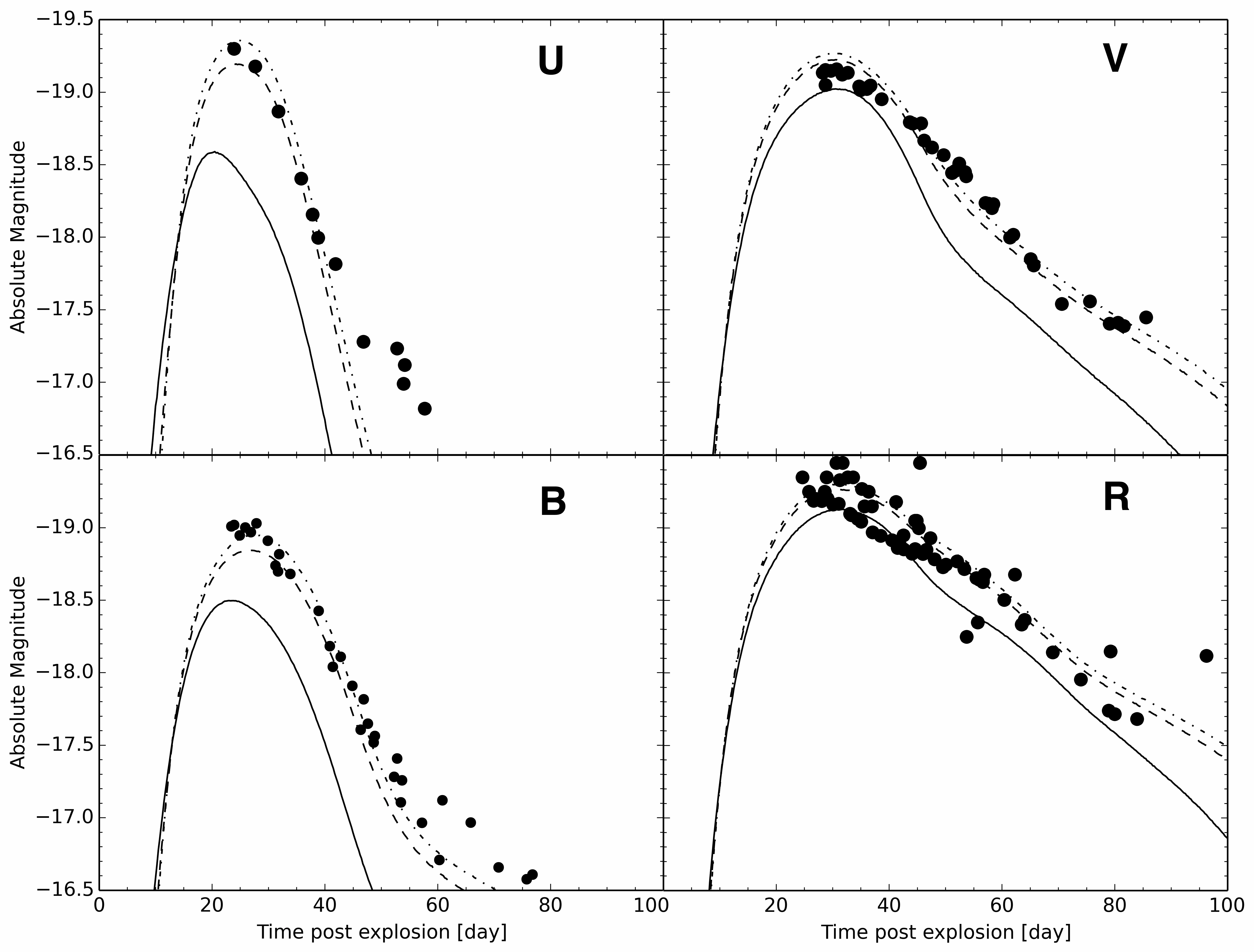}}
\caption{UBVR light curves for three equally spaced polar viewing angle bins with $\mu = \cos(\theta)$   where $\theta$ is the polar angle, and $\mu = 0.89, 0.44 $,and $0$, for the solid, dashed and dot-dashed lines, respectively, compared against data from the slowly declining SN 2001ay (solid points).
% Viewing angle bins cover the range in $\mu$ from -1 (south) to 1 (north) and are regularly spaced in terms of solid angle.
 %In order to simplify the figure, the light curves are averaged over azimuthal viewing angles $\phi$ and the viewing angle bins with $\mu$ of -0.67, -0.22, 0.22, and 0.67 are omitted.
% The variation with azimuthal angle is significantly smaller than the variation with polar angle.
% The light curves reflect the high degree of north-south symmetry in the ejecta as visible in Figures~\ref{fig:particles} and~\ref{fig:particles-cylinder}.
 %The bolometric luminosity peaks around day 30 post explosion, which is later than the typical time to peak of approximately 20 days for normal \sneia.
% This is due to relatively low ejecta velocities and \nifs\ mass compared to the total ejecta mass, which leads to slower expansion, and longer time for the ejecta to become optically thin to the region where \nifs\ decays.
}\label{fig:lc-bol}
\end{figure}

\begin{figure}
\centerline{\includegraphics[width=.5\textwidth]{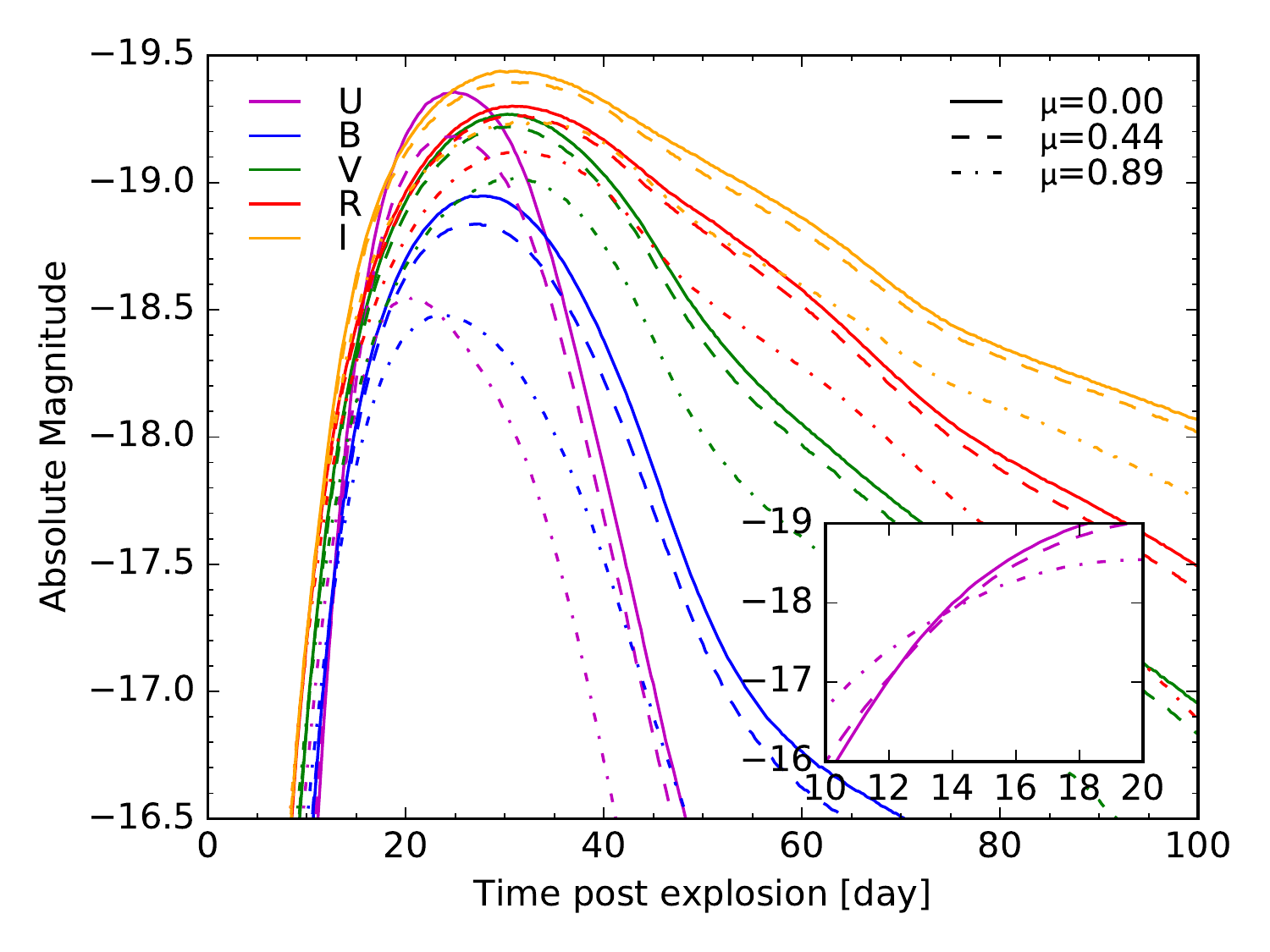}}
\caption{UBVRI-band light curves at three polar viewing angles (from equatorial to north) azimuthally averaged.
 The light curves for the southern viewing angle bins (at $\mu$= -0.44 and -0.89) are nearly identical to those at $\mu$ = 0.44 and 0.89, respectively, and are omitted to simplify the figure.}
 \label{fig:lc-band}
\end{figure}

Figure~\ref{fig:lc-bol} shows the bolometric light curves calculated for the spiral merger simulation with \supernu, and figure~\ref{fig:lc-band} shows the multiband light curves.
The light curves are averaged over the azimuthal viewing angle in order to simplify the figure and because the variation with azimuthal angle is significantly smaller than the variation with polar angle.
Viewing angle bins cover the range in $\mu$ from -1 (south) to 1 (north) and are regularly spaced in terms of solid angle. 
Light curves are plotted for three polar viewing angles: $\mu = 0.89, 0.44 $,and $0$.  
 %In order to simplify the figure, the light curves are averaged over azimuthal viewing angles $\phi$ and the viewing angle bins with $\mu$ of -0.67, -0.22, 0.22, and 0.67 are omitted.

The light curve color (B-V) at peak brightness is higher (redder) than for normal \sneia, which have (B-V)$\approx$0.
 This is due to the relatively low ejecta velocities and \nifs\ mass compared to the total ejecta mass, which leads to lower temperatures at the time of peak.
 
The inset in figure~\ref{fig:lc-band} shows a close-up of the U-band light curves around day 15\,pe (days post-explosion), where the polar viewing angle dependency of the brightness reverses from brighter toward the poles before day 15\,pe to dimmer toward the poles afterward.
This is caused by the asymmetric distribution of \nifs\ in the ejecta.

The agreement with SN 2001ay is excellent across all bands, although the U-band data shows a slight excess over the model at times later than 50 days.
% The inset figure shows a close-up of the U-band light curves around day 15\,pe, where the polar viewing angle dependency of the brightness reverses from brighter toward the poles before day 15\,pe to dimmer toward the poles afterwards.
%This is caused by the asymmetric distribution of \nifs\ in the ejecta

%Multi-band light curves are shown in Figure~\ref{fig:lc-band} for three viewing angles.
%

%

%-----------------------------------------------------------------------
\subsection{Phillips Relation}

\begin{figure}
\centerline{\includegraphics[width=.5\textwidth]{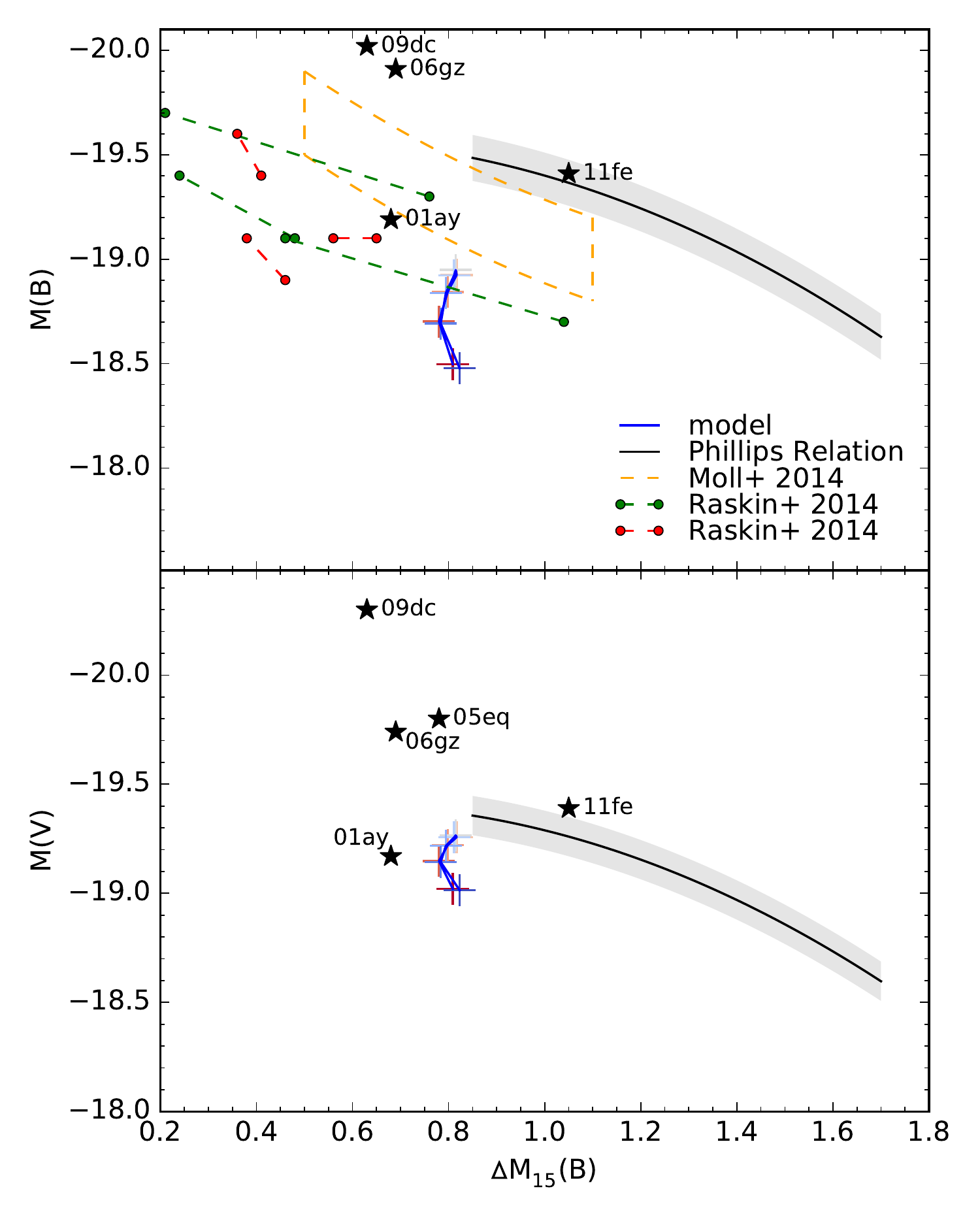}}
\caption{Width-luminosity relation of the spiral merger's synthetic light curves at different polar viewing angles (crosses, blue line) compared to the observational relation from~\citet{Phillips99} in the B and V bands (top and bottom panels).
 Note that the width, \dmft, is measured in the B band in both panels, as usual.
 The color of the crosses changes with polar viewing angle from red (south) to blue (north), with gray being the central polar viewing angle bin that contains the equatorial plane.
 The \dmft\ is smaller than for normal \sneia, such as 2011fe, and as represented by the standard deviation (gray shaded area) of the Phillips relation.
 Results from several other two-dimensional merger models with mass ratios similar to 1.1 or significantly higher are plotted with dashed lines: violent mergers with mass ratio $\approx 1.1$ \citep{Moll14} (orange region), and delayed detonation mergers with mass-ratio $\approx 1.1$ (\citet{Raskin_2014}; green), and high mass-ratio 1.5--2.5 (red).
 The small \dmft\ of the spiral merger and its relatively low brightness are similar to the slowly declining SN 2001ay \citep{krisciunasetal11}.
}\label{fig:phillips}
\end{figure}

Figure~\ref{fig:phillips} plots the width-luminosity relation of the synthetic light curves compared to the Phillips relation \citep{Phillips99}, where the luminosity and width are represented by the peak B-band and V-band magnitude ($M$(B) and $M$(V)) and the decline in B-band magnitude after 15 days from the time of the peak (\dmft).
The shaded area around the analytic Phillips relation highlights the $1\,\sigma$ confidence level in the \dmft\ range over which the Phillips relation was determined \citep{Phillips99} and which approximately describes the range that is covered by normal \sneia.
The peak brightness reached in the spiral merger simulation is compatible, albeit on on the dim end, with the normal brightness range, but at the same time the decline is much slower than that of normal \sneia\ with that brightness.

Figure~\ref{fig:phillips} also compares the spiral merger and violent merger models against several slowly declining SNe Ia, including SNe 2009dc \citep{taubenbergeretal11}, 2006gz \citep{hickenetal07}, and 2001ay \citep{krisciunasetal11} in the B and V bands. SNe 2009dc and 2006gz are overluminous, with a variety of unusual properties including strong unburned C \textsc{ii} and IME lines at early times, and belong to a family of SNe Ia including SN 2003fg, which are interpreted as super-Chandrasekhar-mass systems by some authors \citep{howelletal06}. While the \dmft\ of the spiral merger model is comparable to these overluminous systems, it  is dimmer by 1 -- 1.5 mag. The spiral merger model light curves instead more closely match SN 2001ay, including a remarkably close match in the V band. 

\subsection{Gamma-ray light curves}\label{sec:grlc}

\begin{figure}
\centerline{\includegraphics[width=.5\textwidth]{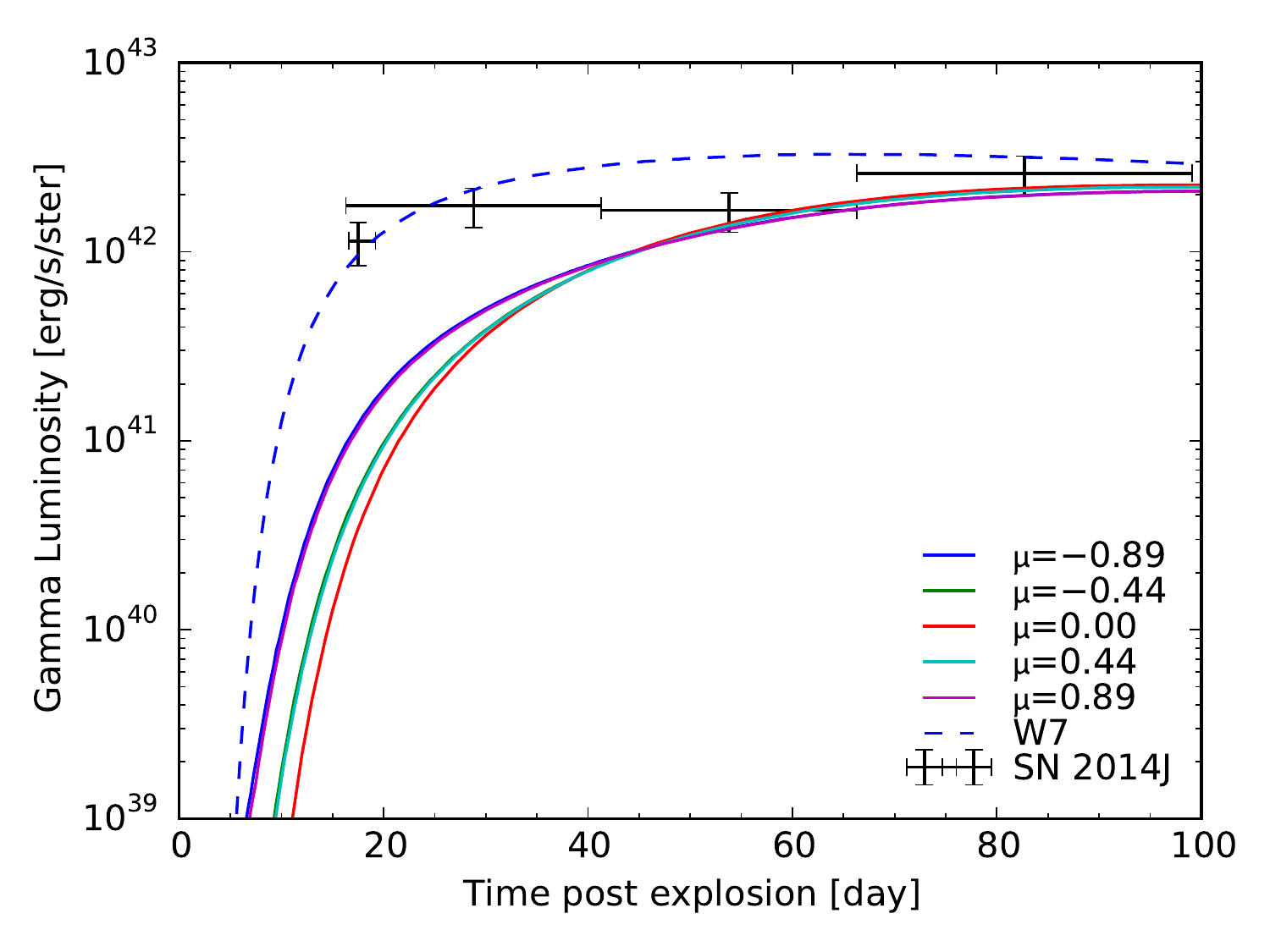}}
\caption{Synthetic gamma-ray light curves for the same polar viewing angle bins plotted in Figure~\ref{fig:lc-bol}, averaged over azimuthal viewing angle and compared to data for SN 2014J (see text) and the W7 model.
 Before day 45\,pe, the gamma-ray luminosity is lowest in the central viewing angle bin and increases toward the poles.
 This is due to the asymmetric distribution of mass and \nifs\ in the ejecta.
 The early gamma-ray flux being higher toward the poles causes the early (before day 15\,pe) optical light curves to be brighter and bluer toward the poles in Figures~\ref{fig:lc-bol} and~\ref{fig:lc-band}.
 The rise in the gamma-ray light curve is slower than observations for SN 2014J, which are better reproduced by the classical W7 model.
}\label{fig:lc-gamma}
\end{figure}

It is interesting to note that the bolometric light curves in Figure~\ref{fig:lc-bol} and the multiband light curves in Figure~\ref{fig:lc-band} are brighter and bluer toward the poles before day 15\,pe but dimmer and redder afterwards.
This is due to the asymmetric distribution of mass and \nifs\ in the ejecta that gives rise to strong polar angle-dependent early gamma-ray flux, as demonstrated in Figure~\ref{fig:lc-gamma}.
The gamma-ray flux heats the outer regions of the ejecta, so that a higher gamma-ray flux leads to higher temperatures which in turn lead to higher optical luminosities and bluer spectra.

The gamma-ray opacity is proportional to the electron density $n_{\rm e} \propto (v t)^{-3}$, where $v$ and $t$ are the expansion velocity and the expansion time.
Let us consider the contributions to the gamma-ray flux in different polar viewing angle bins from \nifs\ that is located either at low velocities and close to the $z=0$ plane, or in the lobes around the $z$-axis at higher velocities.
In early epochs, the gamma-ray optical depth in the $\mu = 0$ bin to both the low-velocity \nifs\ and the high-velocity \nifs\ in the lobes is relatively high because the optical paths to each of these regions pass through low-velocity material with relatively high electron density.
At the poles, the early gamma-ray optical depth to the closer \nifs\ lobe is relatively low because the optical path to that lobe passes through material with high expansion velocities and relatively low electron densities.
Consequently, the early gamma-ray flux is low in the $\mu = 0$ bin and higher toward the poles.
Over time, the gamma-ray optical depths from the $\mu = 0$ bin decrease to the low-velocity \nifs\ \emph{and} to the high-velocity \nifs\ in both lobes, and all three regions will contribute to the flux in this bin.
At the poles, however, while the optical depth to the closer lobe continues to be relatively small, the optical depths to the other two regions, that is, the low-velocity \nifs\ and the \nifs\ in the far lobe, is much higher than from the $\mu = 0$ bin, so that the contributions to the gamma-ray flux from those regions will stay small much longer.
We discuss these findings in connection with recent unexpectedly early gamma-ray observations of the \nifs\ decay lines from 2014J \citep{Diehl14} below in the discussion section, \S\ref{sec:Conclusions}.

Figure~\ref{fig:lc-gamma} compares the angle-dependent synthetic gamma-ray flux to the gamma flux from nearby supernova SN 2014J, as reconstructed from measured gamma line intensities reported by \citet{Diehl14,Diehl15}.
We combine the line intensities, $I$, of the 158 and 812 keV \nifs\ decay lines and the 847 and 1238 keV \cofs\ decay lines, measured in epoch $t$, into a decay rate
\begin{equation} \label{eq:decayrate}
 \dot{n_j}(t) = 4\pi D^2 \frac{1}{n} \sum_{i=1}^n I_{ij}(t) / b_{ij} \;\;,
\end{equation}
where the sum is over $n=2$ lines $i$, $j$ is either \nifs\ or \cofs, and $b_{ij}$ are the branching ratios 1.0 and 0.86 for the two \nifs\ decay lines, and 1.0 and 0.68 for the two \cofs\ decay lines.
$D = 3.27$ Mpc is the assumed distance to SN 2014J \citep{Foley14}.
Using Equation~\ref{eq:decayrate} for either \nifs\ ($j=1$), or \cofs\ ($j=2$) data, the total gamma flux $F_\gamma$ can be calculated as
\begin{eqnarray}
F_\gamma(t) = \frac{\dot{\epsilon}(t)}{\dot{\epsilon_j}(t)} Q_j \; \dot{n_j}(t) \;\;,
\end{eqnarray}
where $\dot{\epsilon} = \dot{\epsilon}_{\rm Ni} + \dot{\epsilon}_{\rm Co}$ is the instantaneous gamma decay energy rate due to \nifs\ and \cofs\ decay at time $t$, and $Q_j = \sum_i Q_{ij}$ is the gamma-ray energy release in all lines per decay of species $j$; $Q_{\rm Ni} = 1.75$ MeV and $Q_{\rm Co} = 3.61$ MeV.
The data point at $t = 17.5$ days post-explosion (pe) comes from the \nifs\ lines and the later data points from the \cofs\ lines.
The synthetic gamma-ray flux from the spiral merger simulation is lower than the early detected flux levels by a factor of 10.
For comparison, Figure~\ref{fig:lc-gamma} also shows the synthetic gamma-ray flux from the classical W7 model \citep{Nomoto84}, which was reported to be in agreement with the observations by \citet{Diehl15}.

%-----------------------------------------------------------------------
\subsection{Synthetic Spectra}

\begin{figure*}
\centerline{\includegraphics[width=\textwidth]{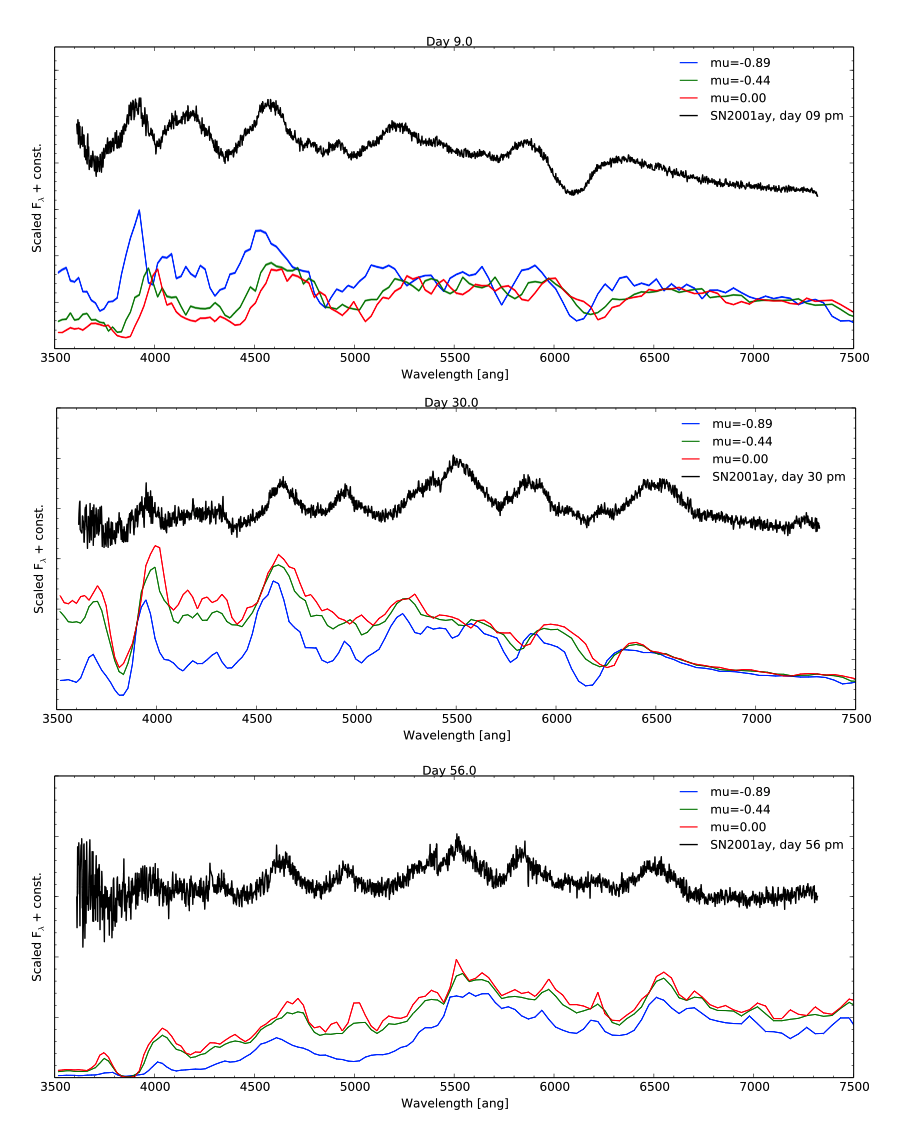}}
%\centerline{\includegraphics[width=\textwidth]{super3d.2r60.d15.Spec.pdf}}
%\centerline{\includegraphics[width=\textwidth]{super3d.2r60.d30.Spec.pdf}}
%\centerline{\includegraphics[width=\textwidth]{super3d.2r60.d60.Spec.pdf}}
\caption{Synthetic spectra at day 9 (top), 30 (middle), and 56 (bottom) pe shown from three different polar viewing angles, averaged over the azimuthal viewing angle, and compared to the vertically shifted spectra for SN 2001ay at the same times post-maximum (pm). 
}\label{fig:spec}
\end{figure*}

Figure~\ref{fig:spec} shows the synthetic spectra of the spiral merger simulation at days 9, 30, and 56 pe compared to SN 2001ay \citep {krisciunasetal11}.
%
%The Hsiao templates are averages over spectra from a heterogeneous set of normal \sneia, each corrected for light curve stretch to match a mean rise and decline rate. In the procedure generating the templates, features that are typical for only fast  \sneia\ or only slow \sneia\ become less pronounced in the mean, but nonetheless common features remain. In order to compare the templates to individual supernovae, or simulations, the templates have to be stretched to match the rise and decline rate of the light curves of that particular object. For the comparison in Figure~\ref{fig:spec} we align the templates at day 0 post maximum with day 30\,pe, and apply a stretch factor of 1.5.

%The pre-maximum model spectra taken at day 15 are lower in the UV\@, compared to the Hsiao spectra.
The variation of the model spectra with viewing angle at all epochs is remarkably small in comparison to the synthetic spectra of violent mergers, with the most significant effect being the amount of blueshift with which the absorption features appear.
%The dispersion with viewing angle increases over time, At the time of maximum brightness at day 30, the dispersion with viewing angle is a bit larger.
The spectra seen from the poles are redder than the spectra viewed at angles closer to the midplane. 
The midplane spectra match the observations fairly well.
This trend with polar viewing angle is the opposite from that seen in the pre-maximum spectra where the polar spectra were brighter in the UV, as expected from Figure~\ref{fig:lc-band}.
This is due to the asymmetric distribution of mass and \nifs\ in the ejecta that gives rise to strong polar angle-dependent early gamma-ray flux, as discussed in Section~\ref{sec:lc}.
At late epochs, well after peak, the model spectra are too red at all viewing angles, but the dispersion with viewing angle is smaller than at peak.

Various absorption troughs of p-Cygni features appear to be less blue-shifted than when viewed from the equator, as expected from the low expansion velocities around the equator.
These include the Si\,\textsc{ii} line around 6150\,\AA\ and the Ca\,\textsc{ii} IR-triplet around 8150\,\AA\ at day 52.5\,pe.
The Ca\,\textsc{ii} HK line around 3750\,\AA\, appears less blue-shifted than normal SNe Ia from all viewing angles.
The sulfur ``w'' feature develops in the spectra of SN 2001ay around 5000\AA by +9 d, and is pronounced at +30 days and +56 days.
In the model spectra, the sulfur feature develops slightly later and is first visible from near the midplane $\mu = -0.89$ spectra at +30 d.
Some key line velocities determined from the spectra are listed in Table~\ref{tab:velline}.

\begin{deluxetable}{lcccc}
\tablecaption{Line Velocities in the Synthetic Spectra in Units of $10^3$ km/s\label{tab:velline}}
\tablehead{\colhead{Day pe} & \colhead{Feature} & \colhead{$\mu=0.89$} & \colhead{$\mu=0.44$} & \colhead{$\mu=0.00$}}\\
\startdata
15.0 & Ca~\textsc{ii} 3750\AA\ & 12.1 & 10.6 & 9.9 \\
15.0 & Ca~\textsc{ii} 8150\AA\ & 10.1  & 9.9 & 7.4\\
15.0 & Si~\textsc{ii} 6150\AA\ & 11.1 & 8.3 & 5.4\\
       
30.0 & Ca~\textsc{ii} 3750\AA\ & 11.3 & 9.9  & 9.4 \\
30.0 & Ca~\textsc{ii} 8150\AA\ & 10.6 & 9.6 & 7.2\\
30.0 & Si~\textsc{ii} 6150\AA\ & 10.6 & 7.3 & 5.0\\
       
52.5 & Ca~\textsc{ii} 3750\AA\ & 6.8 & 7.6 & 7.6 \\
52.5 & Ca~\textsc{ii} 8150\AA\ & 9.2  & 7.4 & 7.4
\enddata
\end{deluxetable}

\begin{figure}
\centerline{\includegraphics[width=.5\textwidth]{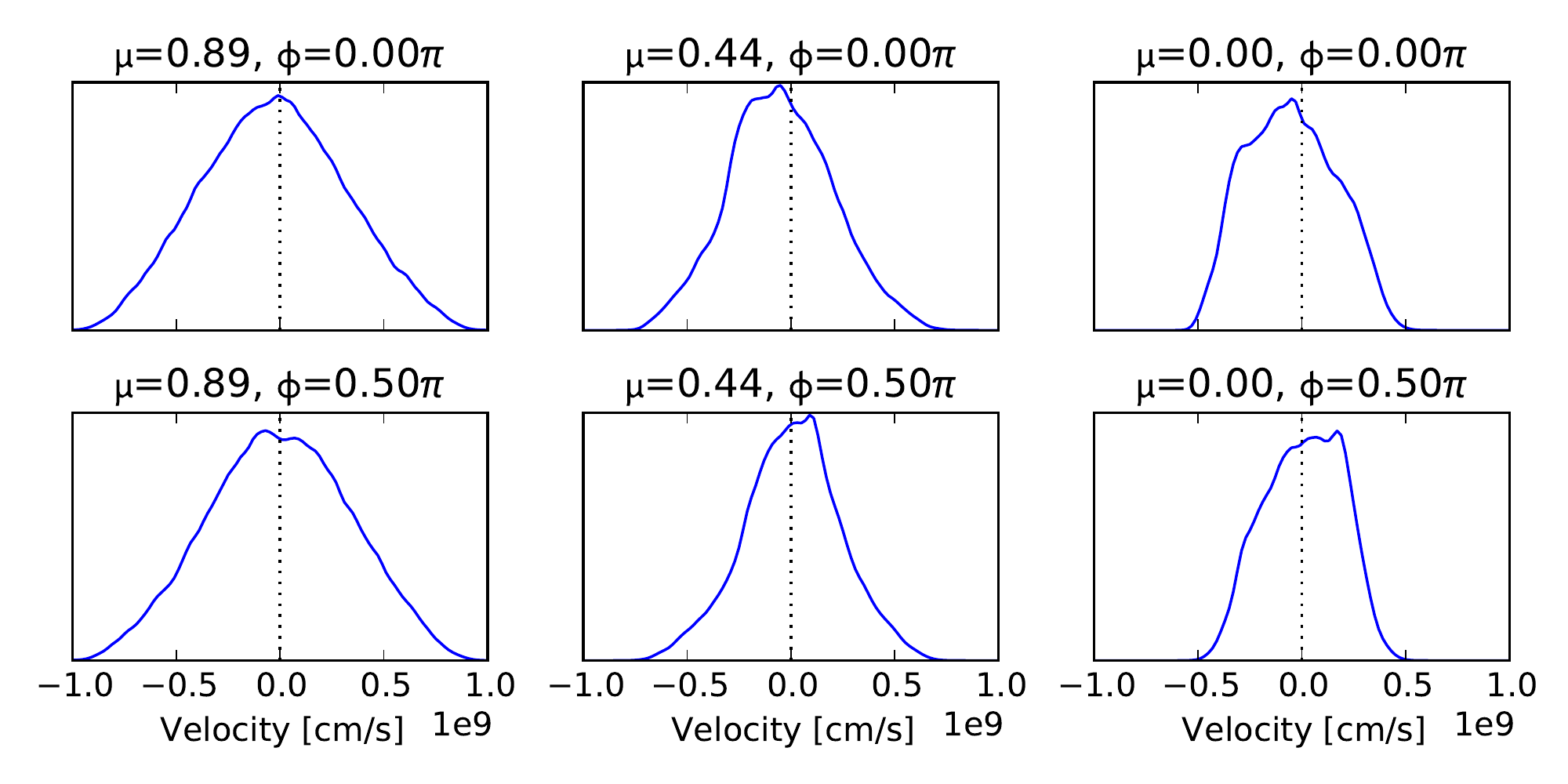}}
\caption{Velocity distribution of \nifs\ along lines of sight with different polar ($\mu = \cos(\theta)$) and azimuthal ($\phi$) viewing angles.
 No clearly doubly peaked distributions are produced in the spiral merger simulation.
}\label{fig:vel-los}
\end{figure}

\citet{Dong15} have reported three SNe Ia that show doubly peaked line profiles of well-separated Co and Fe features in nebular spectra.
Such profiles suggest that in those observations the \nifs\ distribution in the ejecta is bimodal and viewed from a direction aligned with the bimodality.
They demonstrate a DD collision simulation with a small non-zero impact parameter of 0.2 that gives rise to two \nifs\ components that are separated by several thousand km/s and features doubly peaked velocity distributions for certain lines of sight \citep[Figure 5]{Dong15}.
Figure~\ref{fig:vel-los} shows the velocity distributions along six lines of sight for the spiral merger simulation.
Apparently, the hourglass morphology of the \nifs\ in the ejecta is not bimodal enough to produce doubly peaked velocity distributions due to the continuous transition between the northern and the southern \nifs\ lobes. We infer that such doubly peaked Co and Fe lines require a strong degree of asymmetry in the explosion mechanism, greater than that generated in the spiral merger model.

%-----------------------------------------------------------------------
\section{Discussion and Conclusions}\label{sec:Conclusions}

We present the light curves and spectra from a thermonuclear explosion of a WD that initiates self-consistently when at the base of an accretion disk with spiral instability a stream of hot, low-density accretion material mixes with cold, high-density material from the primary.
In the spiral merger simulation, a high degree of reflection symmetry across the plane of the spiral disk at $z=0$ develops.
This leads to light curves and spectra with a modest viewing-angle dependence that is smaller than the viewing angle dependence reported for violent merger and delayed detonation merger simulations \citep{Moll14,Raskin_2014}.
The ejecta reach higher expansion velocities toward the poles than close to the $z=0$ plane because the ejecta transfer radial momentum to the spiral disk during the expansion, and therefore slow down.
This manifests in the synthetic spectra through various absorption features that appear more blue-shifted when viewed from the poles than from the equator.

An interesting consequence of the asymmetric ejecta morphology is that before day 15\,pe, the UV flux is stronger toward the poles than around the equator, but after day 15\,pe it becomes dimmer toward the poles than around the equator.
In early epochs, the angle-dependent gamma-ray flux heats the outer ejecta more toward the poles than toward the disk.
This leads to higher temperatures around the poles and causes stronger UV flux.
At later times, when the ejecta have expanded more and the gamma-ray optical depths have decreased, the ejecta around the $z=0$ plane are heated by gamma-rays from both the northern and the southern \nifs\ lobes, leading to higher temperatures than in the outer regions toward the poles for which the far lobe is in the ``shadow'' of the closer lobe.
This causes the angle dependence in the UV flux to reverse around day 15\,pe.

A comparison of the synthetic gamma-ray light curves to the recent gamma-ray observation of SN 2014J \citep{Diehl14} shows that the rise time of the predicted gamma-ray light curve is much longer than what is observed for this particular supernova.
At day 17.5\,pe, the time of the first gamma-ray measurement for SN 2014J, the discrepancy between observed and synthetic flux is as large as a factor 10, or $3\,\sigma$.
Comparison of the spiral merger simulation with the classical W7 model shows that the gamma-ray light curves are very distinct between the spiral merger and the W7 model.
These two models have approximately the same \nifs\ mass but the total mass in the W7 model is 35\% less.
The gamma-ray measurements from SN 2014J, and in particular the narrow lines observed at early times, therefore provide a unique observational challenge to theory with significant discriminatory power to distinguish between models.

The rise times for the gamma-ray light curves and the optical light curves are governed by the fundamental properties of a system with the super-\mch\ total ejected mass and \nifs\ mass derived from the detonation of the sub-\mch\ CO WD primary in two ways.
First, a thicker blanket covers the nuclear decay energy source so that more expansion is required for the source to heat the outer ejecta, i.e., the regions visible from the outside, and brighten the light curve.
Second, given the large total mass of the system relative to the burned mass, which determines the energy release during the explosion, the ejecta velocities are relatively low.
Therefore, a longer time is required for prompt super-\mch\ DD mergers to reach peak brightness and the systems have lower temperatures at peak brightness, and therefore redder colors.
These effects are generic to any prompt DD system which undergoes a detonation on a dynamical timescale subsequent to merger.

The slow evolution of this spiral merger simulation is in agreement with results published for violent merger and delayed detonation merger simulations \citep{Moll14,Raskin_2014}. 
The spiral detonation arises at a stage of evolution intermediate between the violent mergers considered by \citet{Moll14} and the delayed detonation mergers considered by \citet{Raskin_2014} and consequently have characteristics of both.
We find that our nucleosynthetic yields, particularly the lower abundances of \nifs\ and greater abundances of intermediate-mass isotopes like \site, resemble the prompt mergers considered by \citet{Moll14}, while the hourglass morphology more closely resembles the \citet{Raskin_2014} delayed mergers.
As a result, the spiral merger models exhibit both the spectral lines of intermediate-mass elements characteristic of SNe Ia as well as a reduced viewing angle sensitivity -- both favorable properties in connection to explaining normal SNe Ia.

Furthermore, we find that a family of very slowly declining normal SNe Ia, including SN 2001ay and its close relatives (SN 2005eq, SN 2006gz), provide relatively similar light curves and spectra to the spiral merger model.
Previous attempts to explain SN 2001ay invoked relatively extreme scenarios.
For instance, simple spherical estimates found a total system mass of 4.4 $M_{\odot}$, which would seem to require exotic non-standard super-\mch\ WD binary models \citep{krisciunasetal11}.
Other detailed calculations suggested that SN 2001ay could be accommodated by a \mch\ white WD explosion, but only if the WD had 80\% carbon \citep{baronetal12}. 
The difficulty in accounting for these normal, slowly declining systems in the \mch\ channel and simple spherical DD models is in contrast to the natural way in which they fit into a more realistic spiral merger model.
Consequently, based on the natural similarity of both the slowly declining light curve and the spectral properties of SN 2001ay and its close relatives to spiral mergers, these systems represent the closest correspondence found yet to date between observed normal SNe Ia and DD mergers.
Still, there are some systematic deviations between the spiral merger model and observations of SN 2001ay. In particular, the velocities of the Si\,\textsc{ii} lines in the spiral merger model are greatest along the poles, as also found by \citet{Raskin_2014}, but are nonetheless lower than observed. 
The higher velocities in SN 2001ay  may be the result of a somewhat larger \nifs\ yield and greater kinetic energy than the spiral merger simulation presented here, possibly due to greater accretion from the disk onto the primary.

Looking to the future,  the similarity between the family of normal, slowly declining SNe including SN 2001ay and the relatively rare, massive spiral merger considered here hints at the possibility that such rare DD mergers may be the tip of the iceberg, with more typical mergers accounting for the majority of normal SNe Ia. In particular, more typical, lower-mass DD mergers will naturally be more rapidly declining, and may account for the broader class of normal SNe Ia with more rapidly declining light curves. These lower-mass DD mergers may result from magnetically driven disk accretion in a near-equal mass system \citep {vankerkwijketal10, jietal13}, and a key challenge to theory is the full explication of the mechanism by which such systems may detonate.

\hspace{.5cm}
\acknowledgements{}
We acknowledge useful discussions with Jerod Parrent, Marius Dan, and Brad Schaefer. This work is supported in part at the University of Chicago by the National Science Foundation under grants AST--0909132, PHY--0822648 (JINA, Joint Institute for Nuclear Astrophysics), and PHY--1430152 (JINA-CEE, Joint Institute for Nuclear Astrophysics).
This work used the Extreme Science and Engineering Discovery Environment (XSEDE), which is supported by National Science Foundation grant number ACI-1053575. Simulations at UMass Dartmouth were performed on a computer cluster supported by NSF grant CNS-0959382 and AFOSR DURIP grant FA9550-10-1-0354.
The work of E.G.-B., G.A.-S. and P.L.-A. was partially funded by the MINECO AYA2014-59084-P grant and by the AGAUR\@.
This research has made use of NASA's Astrophysics Data System and the yt astrophysics analysis software suite \citep {turketal11}.

\bibliography{bibDaan,fisher_sne}

\end{document}